\documentclass[twocolumn,showpacs,preprintnumbers,amsmath,amssymb]{revtex4}
\usepackage[dvips]{graphicx}
\usepackage{dcolumn}
\usepackage{bm}
\begin{document}

\title{Noise of a single-electron 
transistor in the regime of large quantum fluctuations 
of island charge out of equilibrium
}

\author{Yasuhiro Utsumi}
\altaffiliation[Present address ]
{Max-Planck-Institut f\"ur Mikrostrukturphysik
Weinberg 2, D-06120 Halle, Germany}

\author{Hiroshi Imamura}
\author{Masahiko Hayashi}
\author{Hiromichi Ebisawa}
\affiliation{Graduate School of Information Sciences, Tohoku University,
Sendai 980-8579, Japan}

\date{\today}
\begin{abstract}
By using the drone-fermion representation
and the Schwinger-Keldysh approach, 
we calculate the current noise and the charge noise for 
a single-electron transistor in the non-equilibrium state 
in the presence of large quantum fluctuation of island charge. 
Our result interpolates between those of 
the \lq \lq orthodox" theory and the \lq \lq co-tunneling theory". 
We find the following effects which 
are not treated by previous theories: 
(i)
At zero temperature $T=0$ and at finite applied bias 
voltage $|eV| \gg T_{\rm K}$, where 
$T_{\rm K}$ is the \lq \lq Kondo temperature", 
we find the Fano factor is suppressed more than 
the suppression caused by Coulomb correlation both in the 
Coulomb blockade regime
and in the sequential tunneling regime. 
(ii)
For $T \gg |eV|/2 \gg T_{\rm K}$, 
the current noise in the presence of large charge fluctuation is modified and
deviates from the prediction of the orthodox theory.
However, the Fano factor coincides with that of the orthodox theory 
and is proportional to the temperature. 
(iii) 
For $eV, T \lesssim T_{\rm K}$, 
the charge noise is suppressed due to the renormalization 
of system parameters caused by quantum fluctuation of charge. 
We interpret it in terms of the modification of the
\lq \lq unit" for island charge. 
\end{abstract}

\maketitle

\newcommand{\rd}{{\rm d}}
\newcommand{\ri}{i}
\newcommand{\mtau}{\mbox{\boldmath$\tau$}}
\newcommand{\mat}[1]{\mbox{\boldmath$#1$}}

\section{introduction}

%
%
In a small metallic island where the charging energy $E_C$ exceeds 
the temperature $T$ (we use the unit $k_{\rm B}=1$), 
the Coulomb interaction affects transport properties through the island. 
The resulting phenomenon is called the Coulomb blockade (CB)
and such system is named the single-electron transistor (SET). 
CB has been attracted much attention in the last decade
\cite{Averin_Likharev,Ingold_Nazarov,Schon_R} and 
the nature of the transport properties of SET has been clarified. 
SET is interesting because it is regarded as one of the most simple examples 
of strongly correlated system which can be brought into the 
non-equilibrium state by applied bias voltage. 
Early investigations consider the case where the tunneling conductance is 
so small that the higher order quantum fluctuation of 
island charge is negligible. 
Recently, the quantum fluctuation in SET has attracted much attention
as one of the basic problems in this field. 
The quantum fluctuation is quantitatively characterized by the 
dimensionless parallel conductance: 
$\alpha_0=R_{\rm K}/((2 \pi)^2 R_{\rm T})$
where  $R_{\rm K}=h/e^2$ is the quantum resistance 
and $R_{\rm T}$ is the parallel tunneling resistance of 
the source and the drain junctions. 
There has been much development on theoretical investigation
in the whole range of $\alpha_0$. 
Especially in the {\it weak tunneling regime} ($\alpha_0<1$), 
the life-time broadening of a charge state level 
is much smaller than the typical level spacing of charge states, 
and thus the effective two-state model, which is equivalent to the 
multichannel anisotropic Kondo model 
in the equilibrium state\cite{Matveev_CB}, 
well describes the low-energy physics. 
With this model, it is predicted that 
the quantum fluctuation of charge causes the renormalization of 
the conductance and the charging energy 
below the \lq \lq Kondo temperature"
$T_{\rm K}=\frac{E_C}{2 \pi} {\rm e}^{-1/(2 \alpha_0)}$ 
\cite{Matveev_CB,Falci_2,Schoeller_Schon,Schoeller_Konig,Konig_text,Golubev_Zaikin}. 
The renormalization of the conductance is confirmed experimentally
as $1/\ln T$ dependence of the conductance peak at low 
temperature\cite{Joyez}. 
It is also predicted that in the non-equilibrium state, 
the dissipative charge fluctuation
causes the life-time broadening of a charge state level and smears the
structures of $I$-$V$ characteristic\cite{Schon_R,Konig_text,Schoeller_text}.

%
%
Though investigations have revealed much about the quantum fluctuation, 
most of them have been limited to averaged quantities. 
In order to understand the nature of the quantum fluctuation, 
investigations on the higher-order correlation function of
fluctuation operators are required. 
A good starting point may be the investigation on the second moment of 
fluctuation operators, i.e., the noise\cite{Blanter}. 
The charge noise and the current noise in the weak tunneling regime 
is also important 
for practical applications, because it determines the 
performance of SET electrometers
\cite{Schoelkopf,KorotkovR,Averin,Johansson}. 

The current noise is defined by the
auto-correlation function of the current fluctuation operator
$\delta \hat{I}(t)=\hat{I}(t)-\langle \hat{I}(t) \rangle$ as
\begin{eqnarray}
S_{I I}(t,t')
&=&
\langle \{ \delta \hat{I}(t), \delta \hat{I}(t') \} \rangle
\label{eqn:SIIdef},
\end{eqnarray}
%
where $\langle \cdots \rangle$ means the statistical average. 
Until recently, investigations on the noise have been done 
using the framework of 
the \lq \lq orthodox" theory\cite{KorotkovR,Johansson,Hershfield3,Korotkov1,Korotkov2}, 
which takes account of the lowest order quantum fluctuation, 
namely the sequential tunneling (ST) process.  
Recently, several authors\cite{Averin,Brink,Sukhorukov} 
discussed the higher order quantum fluctuation in CB regime
within the \lq \lq co-tunneling theory"\cite{Averin_Nazarov}. 
However, there is no approximation covering both ST and CB regime. 
The aim of the present work is to construct a theoretical framework, 
which covers both of these regimes for arbitrary $\alpha_0$ 
and clarify how the quantum fluctuation affects the noise. 

%
The Keldysh formalism\cite{Langreth,Rammer,Chou} has been one of the most 
powerful methods to study the non-equilibrium properties 
of mesoscopic systems. 
However to apply this method to SET in the two-state limit, 
one must overcome a technical difficulty: 
The spin-1/2 operator, which is introduced to restrict charge number states
by the strong
Coulomb interaction, prevents one from utilizing Wick's theorem.
The most successful treatment to overcome this problem 
is given in Ref. \cite{Schoeller_Schon},
in which a formulation of 
perturbative expansion for the reduced density matrix
in the real time domain is developed and
the inelastic resonant tunneling processes is treated. 
The method of Ref. \cite{Schoeller_Schon} enables one to classify various
tunneling processes using diagrammatic techniques, 
and can be also applied to other systems with 
the strong local correlation, such as quantum dot\cite{Konig1}. 
In spite of these successes, it seems to be still difficult
to apply this method for the calculation of higher order 
correlation functions, 
since this method requires to solve a special 
integro-differential equation
even for the calculation of the average 
in the presence of large quantum fluctuation\cite{Schoeller_Schon}. 

%
%
In this paper, we investigate the current noise and the charge noise 
in the regime of large quantum fluctuation of charge out of equilibrium. 
We adopt the Schwinger-Keldysh approach and
the drone-fermion representation 
of the effective spin-1/2 operator\cite{Isawa,Spencer}. 
Schwinger-Keldysh approach enables us to calculate any order moment
systematically by the functional derivative 
technique\cite{Kamenev_1,Gutman_1}
satisfying the charge conservation\cite{Kamenev_1},
and it helps us in manipulating many complicated terms. 
The drone-fermion representation allows us to utilize the 
fermionic Wick's theorem and to take effects of the 
strong correlation into account.
With the help of this technique, we can extensively
take account of the higher order processes of tunneling. 
We will show that our approximation reproduces 
the resonant tunneling approximation (RTA)\cite{Schoeller_Schon} 
as for the average current and the average charge.

The outline of this paper is as follows. 
In Sec. \ref{sec:Keldysh}, we briefly summarize the Keldysh formalism 
and introduce an approximate generating functional. 
We also show that the average and the noise expressions 
can be derived using the functional derivation. 
In Sec. \ref{sec:main}, we actually calculate the average current, 
the average charge, the current noise and the charge noise. 
In Sec. \ref{sec:results} we show numerical results for the noise and 
give some discussions on the non-equilibrium fluctuation, 
the thermal fluctuation 
and the renormalization effect. 
Section \ref{sec:summary} summarizes our results.

\section{Keldysh formalism and generating functional}
\label{sec:Keldysh}

\subsection{Brief introduction of the Keldysh formalism}
\label{subsec:representations}

\begin{figure}[ht]
\includegraphics[width=.7 \columnwidth]{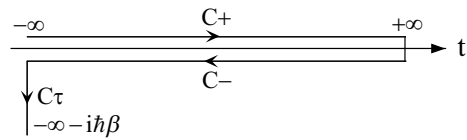}
\caption{
The closed time-path going from $-\infty$ to $\infty$ ($C_{+}$), 
going back to $-\infty$ ($C_{-}$), connecting 
the imaginary time path $C_{\tau}$ and closing at $t=-\infty-\ri \hbar \beta$. 
}
\label{fig:ClosedTimePath}
\end{figure}

In this section, 
we give preliminary definitions of the 
Schwinger-Keldysh approach and 
we summarize three useful representations:
the {\it closed time-path}, 
the {\it single time} and 
the {\it physical} representations
(Sec. 2 of Ref. \cite{Chou}). 
For simplicity, we consider the following action of a free fermion 
with a linear source term to explain the basic formalism: 
\begin{eqnarray}
S
&=&
\int_C \rd t 
\{
a(t)^*
(
\ri \hbar \partial_t
-
\varepsilon
)a(t)
+
a(t) J^{*}(t) + h.c.
\},
\label{eqn:action}
\end{eqnarray}
where the closed time path $C$ 
consists of the forward branch $C_{+}$, the backward branch $C_{-}$
and the imaginary time-path $C_{\tau}$ 
as shown in Fig. \ref{fig:ClosedTimePath}\cite{note10}. 
$a(t)$ is a Grassmann variable satisfying the anti-periodic
boundary condition
$a(-\infty \in C_{+})=-a(-\infty-\ri \hbar \beta \in C_{\tau})$. 
The complex variable $J(t)$
is defined only on the forward and backward branch $C_{+}+C_{-}$. 
The generating functional for the connected closed time-path Green function
(GF) is defined as, 
\begin{equation}
W=- \ri \hbar \ln Z, 
\; \;
Z=
\int {\cal D}
\left[a^*, a \right]
\exp \left(
-S/\ri \hbar
\right). 
\nonumber
\end{equation}
GF in the closed time-path representation
is obtained by the second
derivative of the generating functional with respect to $J(t)$ 
($t \in C_{+}+C_{-}$): 
\begin{equation}
-G(1,2)=
\left.
\frac{\delta^2 W}{\delta J(1)^* \delta J(2)}
\right|_{J=0}.
\end{equation}
Hereafter, we use arguments $1,2$ instead of $t_1,t_2$ for short. 

Though the closed time-path representation makes
the formulation compact, 
in order to obtain the physical quantities, 
we sometimes need the single time representation
in which the time on $C$ is projected onto the real axis. 
In this representation the degrees of freedoms of fields are 
doubled which we denote as, 
\begin{equation}
\hat{\vec{J}}(t)
=
\begin{pmatrix}
J_{+}(t) \\
J_{-}(t)
\end{pmatrix},
\label{eqn:fieldsingletime}
\end{equation}
etc.
Here $J_{\pm}(t)$ is defined on $C_{\pm}$ and $t$ is the real time. 
In the same way as $J$, GF is transformed into 
$2 \times 2$ matrix in the Keldysh space: 
\begin{eqnarray}
\hat{G}(1,2)
=
\begin{pmatrix}
G^{++}(1,2) & G^{+-}(1,2) \\
G^{-+}(1,2) & G^{--}(1,2)
\end{pmatrix}
\label{eqn:CTPGFsingle}.
\end{eqnarray}
Here, arguments $t_1$ and $t_2$ are the real time and each 
component is defined with 
the statistical average in the path integral representation
$\langle A \rangle=
\left.
\int {\cal D} [a^*, a ] A \exp (-S/\ri \hbar)/Z
\right|_{J=0}$
as
$G^{ij}(1,2)=\left. \langle a_{i}(1) a_{j}(2)^* \rangle 
\right|_{J=0}/(\ri \hbar)$. 
Diagonal components $G^{++}$ and $G^{--}$ are 
the causal and the anti-causal GF, respectively. 
Off-diagonal components are correlation functions, 
which are written in the operator representation as
$G^{-+}(1,2)=\langle \hat{a}(1) \hat{a}(2)^{\dagger} \rangle/(\ri \hbar)$
and 
$G^{+-}(1,2)=-\langle \hat{a}(2)^{\dagger} \hat{a}(1) \rangle/(\ri \hbar)$
\cite{Chou}. 
Here the statistical average is defined as 
$\langle \hat{A} \rangle 
= 
{\rm Tr} [ {\rm e}^{-\beta \hat{H}_0} \hat{A} ]/Z_0$ 
where the Hamiltonian operator $\hat{H}_0$ is 
$\varepsilon \hat{a}^{\dagger} \hat{a}$ 
and the partition function $Z_0$ is given as
${\rm Tr} [ {\rm e}^{-\beta \hat{H}_0} ]$. 

It is known that four components of 
Eq. (\ref{eqn:CTPGFsingle}) are not independent. 
This redundancy is removed by the Keldysh rotation\cite{Chou,Rammer}. 
After the rotation, we obtain the so-called physical representation:
\begin{eqnarray}
\tilde{\vec{J}}
=
\left(
\begin{array}{c}
J_1 \\
J_2
\end{array}
\right)
=
\frac{1}{\sqrt{2}}
\left(
\begin{array}{c}
J_{+}-J_{-} \\
J_{+}+J_{-}
\end{array}
\right),
\label{eqn:fieldphysical}
\\
\tilde{G}(1,2)
=
\left(
\begin{array}{cc}
0 & G^A(1,2) \\
G^R(1,2) & G^K(1,2)
\end{array}
\right). 
\end{eqnarray}
GFs denoted by superscripts, $A$, $R$ and $K$ are 
advanced, retarded and Keldysh components, respectively. 
In the practical calculations, instead of $J_1$ and $J_2$, 
the center-of-mass coordinate $J_{c}$ and 
the relative coordinate $J_{\Delta}$,
\begin{eqnarray}
\left \{
\begin{array}{c}
{\displaystyle 
J_{c}(t)
=
\frac{J_{+}(t)+J_{-}(t)}{2}
=J_2(t)/\sqrt{2}
}
\\
{\displaystyle 
J_{\Delta}(t)=
J_{+}(t)-J_{-}(t)
=J_1(t)\sqrt{2}
}
\end{array}
\right.,
\label{eqn:physicalsingle1}
\end{eqnarray}
are used in most cases. 

There are the following relations between components of GF: 
\begin{eqnarray}
G^{--}(1,2)+G^{++}(1,2)
=
G^{-+}(1,2)+G^{+-}(1,2)
\nonumber \\
=
\frac{1}{\ri \hbar}
\langle \{ \hat{a}(1)^{\dag},\hat{a}(2) \} \rangle
=
G^K(1,2),
\label{eqn:anti-comute}
\\
G^R(1,2)-G^A(1,2)
=
G^{-+}(1,2)-G^{+-}(1,2)
\nonumber \\
=
\frac{1}{\ri \hbar}
\langle [ \hat{a}(1),\hat{a}^{\dag}(2) ] \rangle
=
G^C(1,2).
\label{eqn:comute}
\end{eqnarray}
Here we introduce a notation $G^C$ whose physical meaning is
the spectral density in the energy space. 
Equation (\ref{eqn:anti-comute}) is derived from the normalization 
of the step function (see Eq. (2.67) in Ref. \cite{Chou}). 

The normalization condition of the density matrix 
results in an important equation
$\left. Z \right|_{J=0}/Z_0=1$
(see Sec. 2.4 of Ref. \cite{Chou})\cite{note1}
which is equivalent to the following equations: 
\begin{eqnarray}
\left. \frac{\delta W}
{\delta J_c(1)} 
\right|_{J_{\Delta}=0}
=
\left. \frac{\delta^2 W}
{\delta J_c(1) \delta J_c(2)} 
\right|_{J_{\Delta}=0}
=\cdots=0. 
\label{eqn:normalization}
\end{eqnarray}

\subsection{Model Hamiltonian in the drone-fermion representation
and the generating functional 
in closed time path - path integral representation}
\label{subsec:W}

In this section, 
we introduce our model Hamiltonian and 
derive the generating functional for SET,
based on which we construct the perturbation theory. 

%
\begin{figure}[ht]
\includegraphics[width=.5 \columnwidth]{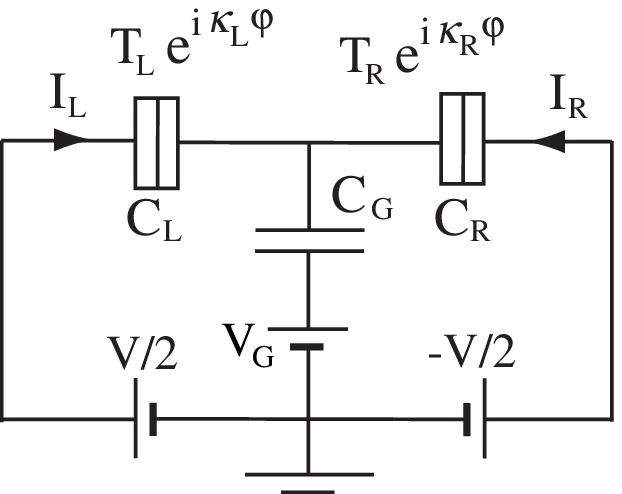}
\caption{
The equivalent circuit of a SET. 
}
\label{fig:system}
\end{figure}
Figure \ref{fig:system} shows the equivalent circuit of a SET.
A metallic island exchanges electrons with the left (right) lead via a tunnel
junction characterized by the tunneling matrix element $T_{\rm L(R)} $. 
The island is coupled to leads and a gate via 
capacitors $C_{\rm L}$, $C_{\rm R}$ and $C_{\rm G}$.
We consider in the weak tunneling regime $\alpha_0 < 1$ and use the 
two-state model. 
In this paper, we limit ourselves to 
the symmetric case: $C_{\rm L}=C_{\rm R}$ and $T_{\rm L}=T_{\rm R}$. 
The total Hamiltonian consists of the unperturbed part $\hat{H}_0$ 
and the tunneling part $\hat{H}_{\rm T}$.
Omitting a trivial constant, the unperturbed part is given as
\begin{equation}
\hat{H}_0
=
\sum_{{\rm r=L,R,I}}
\sum_{k, n}
\varepsilon_{{\rm r} k}
\:
\hat{a}_{{\rm r} k n}^{\dag} \hat{a}_{{\rm r} k n}
+
\Delta_0 \frac{\hat{\sigma}_z+1}{2}, 
\end{equation}
where $\hat{a}_{{\rm r} k n}$ is the annihilation operator of an electron 
with wave vector $k$ in the left (right) lead (${\rm r=L (R)}$) or 
in the island (${\rm r=I}$). 
The subscript $n$ numbers the transverse channels including spin
degree of freedom. 
The density of states is considered as constant in each region: 
$\rho_{\rm r}(\varepsilon)
=\sum_k \delta(\varepsilon-\varepsilon_{{\rm r} \, k})
=\rho_{\rm r}$ ($\rm r=L,R,I$). 
The second term is the charging energy and the effective spin-1/2 
operator $\hat{\sigma}$ acts on the lowest two charge 
states. 
The energy difference between two charge states is given by
$\Delta_0=E_C(1-2 Q_{\rm G}/e)$
where $Q_{\rm G}$ is the gate charge. 

The tunneling Hamiltonian
\begin{equation}
\hat{H}_{\rm T}(t)
=
\sum_{\stackrel{\scriptstyle{\rm r=L,R}}{k, k', n}}
T_{\rm r} e^{\ri \kappa_{\rm r} \varphi(t)}
\:
\hat{a}_{{\rm I} k n}^{\dag} 
\hat{a}_{{\rm r} k' n}
\hat{\sigma}_{+}
+
{\rm h. c.},
\label{eqn:tunnelHamiltonian}
\end{equation}
describes the electron tunneling across the junctions and simultaneous change of the charge state of the island. 
$\varphi(t)=eVt/\hbar$ is the phase difference between the left and 
the right leads and parameters 
$\kappa_{\rm L}=-\kappa_{\rm R}=1/2$ 
characterizes the voltage drop between the left (right) lead and the island. 
The tunneling Hamiltonian is adiabatically turned on 
in the remote past and off in the distant future. 
It is the widely adopted procedure, which ensures 
the time translational invariance 
to describe a stationary state.

In order to utilize Wick's theorem for fermions, 
we employ the mapping of the effective spin-1/2 operator 
onto two fermion operators 
$\hat{c}$ and $\hat{d}$ \cite{Isawa,Spencer,Mattis}:
$
\hat{\sigma}_{+}=\hat{c}^{\dag} \hat{\phi}, 
\:
\hat{\sigma}_{z}=2 \hat{c}^{\dag} \hat{c}-1
$, 
where $\hat{\phi}=\hat{d}^{\dag}+\hat{d}$ is a Majorana fermion operator
($\hat{\phi}^2=1$). 
This representation is called drone-fermion representation
\cite{Spencer}, because $\hat{\phi}$ is a \lq \lq drone" whose
only job is to make spin-1/2 operators of different spins commute, 
rather than anti-commute\cite{Mattis}.

Employing the Hamiltonian operator in 
the drone-fermion representation 
and following the standard manner to introduce a path integral
\cite{Babichenko}, 
we obtain the generating functional in the path integral representation: 
\begin{eqnarray}
Z
&=&
\int {\cal D}
\left[ a_{{\rm r} k n}^*,a_{{\rm r} k n},c^*,c,d^*,d \right]
\exp \left(
-\frac{S}{\ri \hbar}
\right), 
\label{eqn:Z}
\end{eqnarray}
where all field variables are Grassmann variables satisfying the 
anti-periodic boundary condition. 
The action is given by
\begin{eqnarray}
S
&=&
\int_C \rd t 
\{
c(t)^* (\ri \hbar \, \partial_t-h(t)) c(t)
+
\ri \hbar \, d(t)^* \partial_t d(t)
\nonumber
\\
&+&
\sum_{{\rm r},k, n}
a_{{\rm r} k n}(t)^* (\ri \hbar \, \partial_t-\varepsilon_{{\rm r} k}) a_{{\rm r} k n}(t)
\nonumber
\\
&+&
\sum_{\stackrel{\scriptstyle{\rm r=L,R}}{k, k', n}}
T_{\rm r}
{\rm e}^{\ri \varphi_{\rm r}(t)}
a_{{\rm r} k n}(t)^*
a_{{\rm I} k' n}(t)
\sigma_{+}(t)
+{\rm h. c.}
\},
\label{eqn:action0}
\end{eqnarray}
where, $\varphi_{\rm r}(t)=\kappa_{\rm r} \varphi(t)$. 
In Eq. (\ref{eqn:action0}) 
we introduced auxiliary source fields, 
$h(t)$ and $\varphi(t)$, 
in order to calculate the average and the noise 
by the functional derivation. 
It is noticed that the degrees of freedoms are doubled, 
as shown in Eq. (\ref{eqn:fieldsingletime}).
After the derivation, these variables are put as
$h_{\pm}(t)=\Delta_0$ and $\varphi_{\pm}(t)=eV t/\hbar$ 
to be related with the 
parameters of the actual system. 

By introducing a linear source term $\int_C \rd 1 J(1) \phi(1)$, 
where $J$ is a Grassmann variable, 
all fields can be traced out\cite{note7}. 
In the limit of large transverse-channel number, 
$Z$ is expressed as\cite{Utsumi_3}
\begin{eqnarray}
Z
&=&
\exp
\left(
-\sum_n
\frac{(\ri \hbar)^{2 n}}{n}
{\rm Tr}
\left[
\left(
g_c
\frac{\delta}{\delta J} 
\alpha
\frac{\delta}{\delta J}
\right)^n
\right]
\right)
\nonumber
\\
&\times&
\left.
\exp
\left(
-\frac{1}{2 \ri \hbar}
\int_C \rd 1 \rd 2
\,
J(1)
g_{\phi}(1,2)
J(2)
\right)
\right|_{J=0}
\nonumber \\
&\times&
2 {\rm e}^{{\rm Tr} \left[ \ln g_c^{-1} \right]}, 
\label{eqn:startZ}
\end{eqnarray}
where we omitted the partition function of noninteracting electrons. 
The trace ${\rm Tr}$ and the products represent the 
integration along $C$ as follows, 
\begin{equation}
{\rm Tr} \left[ g_c J g_{\phi} \right]
=
\int_C \rd 1 \rd 2 \, g_c(1,2) J(2) g_{\phi}(2,1).
\nonumber
\end{equation}
%
The particle-hole GF $\alpha=\sum_{\rm r=L,R}\alpha_{\rm r}$, 
in the closed time-path form is written as
\begin{equation}
\alpha_{\rm r}(1,2)
=
- \ri \hbar
\; N_{\rm ch} \; T_{\rm r}^2 g_{\rm r}(1,2) g_{\rm I}(2,1)
\; {\rm e}^{\ri (\varphi_{\rm r}(1)-\varphi_{\rm r}(2))},
\label{eqn:particlehole}
\end{equation}
where $N_{\rm ch}$ is the number of the transverse channels. 
The GF for free electron in the lead 
${\rm r \, (=L,R)}$ and the island ${\rm r \, (=I)}$, 
$g_{\rm r}(t,t')=\sum_{k}g_{{\rm r} \, k}(t,t')$,
is given as
\begin{eqnarray}
g_{{\rm r} k n}^{-1}(t,t')
&=&
(\ri \hbar \, \partial_t - \varepsilon_{{\rm r} k}) \delta(t,t'),
\label{eqn:GFl}
\end{eqnarray}
which satisfies the anti-periodic boundary condition:
$g_{{\rm r} k n}(t,-\infty \in C_{+})
=
-g_{{\rm r} k n}(t,-\ri \hbar \beta-\infty)$. 
The $c$-field and the $d$-field GFs defined by
\begin{eqnarray}
g_c^{-1}(t,t')&=&
(\ri \hbar \, \partial_t -h(t)) \, \delta(t,t'),
\label{eqn:GFc}
\\
g_{\phi}^{-1}(t,t')&=&
\ri \hbar \, \partial_t \delta(t,t')/2,
\label{eqn:GFd}
\end{eqnarray}
also satisfy the anti-periodic boundary condition.

\begin{figure}[ht]
\includegraphics[width=.8 \columnwidth]{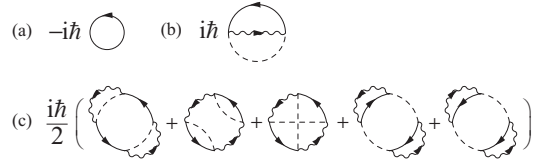}
\caption{
The diagrammatic representation of (a) $W^{(0)}$, (b) $W^{(1)}$ and
(c) $W^{(2)}$. 
Solid lines, dotted lines and wavy lines represent 
GFs for $c$-field,
$d$-field and particle-hole in the closed time-path representation,
respectively.
}
\label{fig:diagram1}
\end{figure}

Using Eq. (\ref{eqn:startZ}), we construct a 
systematic perturbation expansion in terms of $\alpha$. 
For example, we show the diagrammatic representation of 
the zeroth ($W^{(0)}$), 
the first ($W^{(1)}$) and the second order contribution ($W^{(2)}$) 
to the generating functional $W$ 
in Fig. \ref{fig:diagram1}. 
Here, solid lines, dotted lines and wavy lines represent GFs for $c$-field, 
$d$-field and particle-hole, respectively. 
Practical forms are given as follows:
\begin{eqnarray}
W^{(0)}
&=&
-\ri \hbar \,{\rm Tr} \left[ \ln g_c^{-1} \right]
\label{eqn:W0},
\\
W^{(1)}
&=&
\ri \hbar \, {\rm Tr} \left[ g_c \Sigma_c \right]
\label{eqn:W1},
\\
W^{(2)}_{\text{$c$-field}}
&=&
\frac{\ri \hbar}{2}
{\rm Tr}
\left[(g_c\Sigma_c)^2
\right]
\label{eqn:W2c},
\end{eqnarray}
where $W^{(2)}_{\text{$c$-field}}$ is the term 
corresponding to the first diagram in Fig. \ref{fig:diagram1} (c), 
which we call the {\it $c$-field correction}. 
The other four diagrams can be written in the same way. 
Here, the self-energy of the $c$-field 
$
\Sigma_c(1,2)
=
\sum_{\rm r=L,R}
\Sigma_{\rm r}(1,2)
$
is defined as 
\begin{equation}
\Sigma_{\rm r}(1,2)=
-\ri \hbar \, \alpha_{\rm r}(1,2) \, g_{\phi}(2,1).
\label{eqn:partselfenergy}
\end{equation}

\begin{figure}[ht]
\includegraphics[width=.9 \columnwidth]{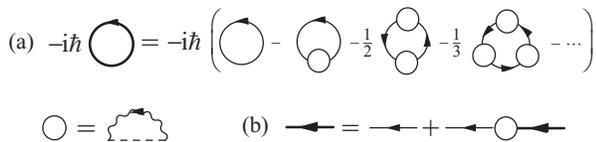}
\caption{
(a) An approximate generating functional including infinite 
order $c$-field correction. 
The circle is the self-energy of the $c$-field. 
(b) The Dyson equation for full $c$-field GF 
in the closed time-path representation. 
}
\label{fig:diagram2}
\end{figure}

In a previous paper\cite{Utsumi_3}, 
we have shown that the first order contribution
causes the divergence at the degeneracy point $\Delta_0=0$ 
for average charge. 
In order to regularize the divergence, we proposed an 
approximate generating functional obtained by summing up 
$c$-field corrections $(g_c \Sigma_c)^n$
to infinite order: 
\begin{eqnarray}
\bar{W}
&=&
- \ri \hbar\,{\rm Tr} \left[\ln G_c^{-1} \right]
\label{eqn:WRTA} \\
&=&
- \ri \hbar
\left(
{\rm Tr}\left[g_c^{-1}\right]
-
\sum_{n=1}^{\infty}
\frac{1}{n}
{\rm Tr} \left[(g_c \Sigma_c)^n \right]
\right).
\nonumber
\end{eqnarray}
Figure \ref{fig:diagram2} (a) shows the diagrammatic representation
of $\bar{W}$. 
The circle represents the self-energy of $c$-field 
and the thick line represents the full $c$-field GF defined by the 
Dyson equation 
\begin{equation}
G_c^{-1}(t,t') = g_c^{-1}(t,t')-\Sigma_c(t,t'),
\label{eqn:fullgc}
\end{equation}
whose diagrammatic representation is shown in 
Fig. \ref{fig:diagram2} (b). 
In Sec. \ref{sec:reformulation} we will show 
that $\bar{W}$ reproduces results of RTA for
the normal metal island. 
In Sec. \ref{sec:mainbody}, 
we calculate the charge noise and the current noise 
based on $\bar{W}$. 

\subsection{Formally exact expressions for average and noise 
and the charge conservation}
\label{sec:moments}

In this section, we summarize expressions for the average and the noise
on the basis of functional derivative. 
Relations between physical quantities 
in the generating functional representation and 
those in the operator representation 
are demonstrated in Appendix \ref{appendix:expressions}.
We also show that the gauge invariance of 
generating functional leads to the charge conservation law. 

The exact average current expression is given by 
the functional derivative of exact generating functional 
$W$ with respect to the phase difference\cite{Kamenev_1,Gutman_1}: 
\begin{eqnarray}
I(t)
=
\left.
\frac{e}{\hbar}
\frac{\delta W}
{\delta \varphi_{\Delta}(t)}
\right|_{
\stackrel{\scriptstyle 
\varphi_c(t)=eVt/\hbar,
h_c(t)=\Delta_0}
{\varphi_{\Delta}=h_{\Delta}=0}
}
\label{eqn:I-exact}.
\end{eqnarray}
The center-of-mass coordinate of the phase difference is determined by
the Josephson relation\cite{Zagoskin,note8}.
The relative coordinates are put to zero because they 
are the fictitious variables. 
From now on, we suppress the equations in the subscript after the 
vertical bar for short.
The average charge is calculated by the functional derivation in terms of 
the scalar potential of $c$-field as
\begin{eqnarray}
\frac{
Q(t)
}{e}
=
\frac{1}{2}
-
\left.
\frac{\delta W}{\delta \, h_{\Delta}(t)}
\right|
\label{eqn:Q-exact}.
\end{eqnarray}

The current noise and the charge noise
defined as Eq. (\ref{eqn:SIIdef}) are given by 
the second derivative with respect to $\varphi_{\Delta}$ and $h_{\Delta}$
\cite{note9}: 
\begin{eqnarray}
S_{I I}(t,t')=
\left. \frac{2 e^2}{\ri \hbar}
\frac{\delta^2 W}{\delta \varphi_{\Delta}(t) \delta \varphi_{\Delta}(t')}
\right|-\frac{(\Delta \rightarrow c)}{4},
\label{eqn:SII}
\\
S_{Q Q}(t,t')=
2 e^2 \left.
\frac{-\ri \hbar \, \delta^2 W}{\delta h_{\Delta}(t) \delta h_{\Delta}(t')}
\right|-\frac{(\Delta \rightarrow c)}{4}
\label{eqn:SQQ},
\end{eqnarray}
where $(\Delta \rightarrow c)$ is obtained from the first term by replacing
subscripts $\Delta$ with $c$. 

It is convenient to rewrite Eq. (\ref{eqn:I-exact})
in the form of the linear combination of the 
tunneling current at the left junction $I_{\rm L}$ and 
that at the right junction $I_{\rm R}$ as 
$
I(t)=\sum_{\rm r=L,R} \kappa_{\rm r} I_{\rm r}(t)
$. 
In the same way, Eq. (\ref{eqn:SII}) is 
written in terms of the correlation function of 
$\hat{I}_{\rm r}$ and $\hat{I}_{\rm r'}$ ($\rm r,r'=L,R$), 
which we denote by
$S_{I{\rm r} \, I{\rm r'}}$, 
as 
$
S_{I I}(t,t')=\sum_{\rm r,r'=L,R}
\kappa_{\rm r} \, \kappa_{\rm r'} \, S_{I{\rm r} \, I{\rm r'}}(t,t'). 
$
Here, $I_{\rm r}$ and $S_{I{\rm r} \, I{\rm r'}}$ are written as
\begin{eqnarray}
I_{\rm r}(t)
&=&
\frac{e}{\hbar}
\left. 
\frac{\delta W}{\delta \varphi_{{\rm r} \, \Delta}(t)} 
\right|
\label{eqn:Ir},
\\
S_{I{\rm r} \, I{\rm r'}}(t,t')
&=&
\left.
\frac{2 e^2}{\ri \hbar}
\frac{\delta^2 W}
{\delta \varphi_{{\rm r} \, \Delta}(t) \delta \varphi_{{\rm r'} \, \Delta}(t')}
\right|
-
\frac{(\Delta \rightarrow c)}{4}
\label{eqn:SIIpart}, 
\nonumber \\
\end{eqnarray}
by regarding $\varphi_{\rm L}$ and $\varphi_{\rm R}$ 
as formally independent variables. 

The generating functional Eq. (\ref{eqn:Z}) is 
invariant under the gauge transformation, 
i.e., the phase transformation of the $c$-field and 
the change of the $c$-field scalar potential:
\begin{eqnarray}
\left \{
\begin{array}{ccc}
\varphi_{\rm r}(t)
& \rightarrow &
\varphi_{\rm r}(t)
+
\delta \psi(t)
\\
h(t)
& \rightarrow &
h(t)
-
\hbar \, \delta (\partial_t \psi(t))
\end{array}
\right.
,
\label{eqn:gauge_t}
\end{eqnarray}
where $\delta \psi$ is defined on $C_{+}+C_{-}$. 
The relation between the gauge invariance and the charge conservation
in the non-equilibrium state has been analyzed in Ref. \cite{Kamenev_1}. 
For our system, the following expressions
of the current continuity and 
the charge conservation for correlation functions 
\begin{eqnarray}
\partial_t Q(t)&=&\sum_{\rm r=L,R} I_{\rm r}(t),
\label{eqn:Iconserve}
\\
\partial_t 
\,
\partial_{t'} 
\,
S_{QQ}(t,t')
&=&
\sum_{\rm r,r'=L,R} S_{I{\rm r} \, I{\rm r'}}(t,t'),
\label{eqn:Sconserve}
\end{eqnarray}
can be proved (Appendix. \ref{appendix:conservationlaw}). 

\section{approximate expressions for noise}
\label{sec:main}

In this section, we derive approximate current noise and charge noise 
expressions, which is main purpose of this paper. 
We summarize GFs in the physical representation 
in Sec. \ref{sec:freeGreenfunctions} 
which are needed for practical calculations. 
In Sec. \ref{sec:reformulation} we calculate the 
average charge and the average current 
and show that our approximation completely reproduces the results of RTA. 
We also give some notes on the diagrammatic rule 
suitable for the functional derivative technique. 
In Sec. \ref{sec:mainbody}, we calculate the 
current noise and the charge noise based on the diagrammatic rule.
We will check
that our results satisfy the charge conservation law
and the fluctuation-dissipation theorem. 

\subsection{Fourier transformation of free Green functions}
\label{sec:freeGreenfunctions}

We summarize the Fourier transformation of 
the retarded and the Keldysh component of GFs. 
Hereafter in this subsection, we put the auxiliary source fields as shown in
the subscript of Eq. (\ref{eqn:I-exact}).
The solutions of differential equations (\ref{eqn:GFc}) and (\ref{eqn:GFd}) 
imposing anti-periodic boundary condition are\cite{note7}
\begin{equation}
g^R_{\phi}(\varepsilon) = 2/(\varepsilon+\ri \eta), \, \,
g^K_{\phi}(\varepsilon) = 0,
\label{eqn:dRK}
\end{equation}
\begin{eqnarray}
\left \{
\begin{array}{ccc}
g^R_c(\varepsilon) &=& 1/(\varepsilon+\ri \eta-\Delta_0)
\\
\displaystyle 
g^K_c(\varepsilon) &=& -2 \ri \pi
\tanh \left( \frac{\varepsilon}{2 T} \right)
\delta (\varepsilon-\Delta_0)
\end{array}
\right.
\label{eqn:cRK}, 
\end{eqnarray}
where $\eta$ is a positive infinitesimal number and 
the $\delta$-function in the energy space is defined as
\begin{eqnarray}
\delta(\varepsilon)=\eta/(\pi (\varepsilon^2+\eta^2)).
\label{eqn:deltae}
\end{eqnarray}
The advanced component is the complex conjugate 
of the retarded component:
$g_{\phi(c)}^A(\varepsilon)=g_{\phi(c)}^R(\varepsilon)^*$.

The two components of particle-hole GF are given by,
\begin{eqnarray}
\left \{
\begin{array}{ccc}
\displaystyle 
\alpha^R_{\rm r}(\varepsilon)
&=&
\displaystyle 
-\ri \pi
\alpha^0_{\rm r}
\rho(\varepsilon-\mu_{\rm r})
\\
\alpha^K_{\rm r}(\varepsilon)
&=&
- 2 \ri \pi
\alpha^0_{\rm r}
\rho(\varepsilon-\mu_{\rm r})
\coth
\left(
\frac{\varepsilon-\mu_{\rm r}}{2T}
\right)
\end{array}
\right.
\label{eqn:alphaRK}, 
\end{eqnarray}
where $\mu_{\rm r}=\kappa_{\rm r} eV$  (Appendix. \ref{appendix:loop}). 
$\alpha^0_{\rm r}$ is the dimensionless conductance for tunnel 
junction ${\rm r}$ written in terms of the tunnel resistance $R_{\rm r}$
as 
$\alpha_{\rm r}^0=
R_{\rm K}/((2 \pi)^2 R_{\rm r})=
N_{\rm ch} T_{\rm r}^2 \rho_{\rm I} \rho_{\rm r}$.
The spectral density of the particle-hole propagator is given by
$\rho(\varepsilon)=\varepsilon E_C^2/(\varepsilon^2+E_C^2)$
where the Lorentzian cut-off function is introduced\cite{Schoeller_Schon}. 
The Keldysh component and the component defined by Eq. (\ref{eqn:comute})
for the $c$-field self-energy Eq. (\ref{eqn:partselfenergy}) 
and those for the particle-hole GF 
are related to each other:
\begin{eqnarray}
\Sigma_{\rm r}^K(\varepsilon)
=
\alpha_{\rm r}^C(\varepsilon), 
\; \; 
\Sigma_{\rm r}^C(\varepsilon)
=
\alpha_{\rm r}^K(\varepsilon).
\label{eqn:selfenergy}
\end{eqnarray}
The retarded component of the self-energy is given as 
\begin{eqnarray}
\Sigma_{\rm r}^R(\varepsilon)
=
\alpha_0^{\rm r}
\rho(\varepsilon)
\left\{
2{\rm Re} \, \psi \left( \ri \frac{\varepsilon-\mu_{\rm r}}{2 \pi T} \right)
\right.
\nonumber \\
-
\left.
\psi \left( 1+\frac{E_C}{2 \pi T} \right)
-
\psi \left( \frac{E_C}{2 \pi T} \right)
\right\}
+
\frac{\alpha_{\rm r}^K(\varepsilon)}{2},
\label{eqn:selfenergyR}
\end{eqnarray}
where $\psi$ is the digamma function. 
The full $c$-field GF is obtained by
solving the Dyson equation in the closed time-path 
representation Eq. (\ref{eqn:fullgc})\cite{Utsumi_3}:
\begin{eqnarray}
& &
\left \{
\begin{array}{ccc}
G_c^R(\varepsilon)
&=&
1/(\varepsilon+\ri \eta-\Delta_0-\Sigma_c^R(\varepsilon))
\\
G_c^K(\varepsilon)
&=&
G_c^R(\varepsilon)
\left \{
\Sigma_c^K(\varepsilon)
-2 \ri \, \eta \tanh \left( \frac{\varepsilon}{2 T} \right)
\right \}
G_c^A(\varepsilon)
\end{array}
\right.,
\nonumber \\
\label{eqn:cG}
\end{eqnarray}
where we used the definition of $\delta$-function 
Eq. (\ref{eqn:deltae}). 
Here we remark the following: 
Equation (\ref{eqn:cG}) shows that 
in the limit of $\eta \rightarrow 0$, 
the charge states are independent of the initial equilibrium 
distribution, 
because the Keldysh component of $c$-field GF 
represents the distribution of charge states. 
This fact suggests that $\bar{W}$ describes 
a physically reasonable non-equilibrium stationary state, 
which should not depend on any initial state. 

We transform above expressions into single-time representation. 
Employing Eqs. (\ref{eqn:anti-comute}) and (\ref{eqn:comute})
we obtain
\begin{eqnarray}
\alpha_{\rm r}^{\pm \mp}(\varepsilon)
&=&
-2 \ri \pi \alpha_0^{\rm r}
\,
\rho(\varepsilon-\mu_{\rm r})
\,
n^{\mp}_{\rm r}(\varepsilon)
=
\mp
\alpha_{\rm r}^K(\varepsilon)
f^{\mp}_{\rm r}(\varepsilon),
\nonumber \\
\label{eqn:correlationalpha}
\\
\Sigma_{\rm r}^{\pm \mp}(\varepsilon)
&=&
\mp
\alpha_{\rm r}^{\pm \mp}(\varepsilon),
\label{eqn:correlationself}
\\
G_c^{\pm \mp}(\varepsilon)
&=&
\mp |G_c^R(\varepsilon)|^2 \alpha^{\pm \mp}(\varepsilon)
\label{eqn:correlatonfullgc1}.
\end{eqnarray}
Here
$f_{\rm r}^{-}(\varepsilon)=f^{-}(\varepsilon-\mu_{\rm r})$ 
and
$n_{\rm r}^{-}(\varepsilon)=n^{-}(\varepsilon-\mu_{\rm r})$ 
are written with 
the Fermi function 
$f^{-}(\varepsilon)=1/({\rm e}^{\beta \varepsilon}+1)$ 
and 
the Bose distribution function
$n^{-}(\varepsilon)=1/({\rm e}^{\beta \varepsilon}-1)$. 
Functions
$f^{+}$ and $n^{+}$ are given by 
$
f^{+}(\varepsilon)=
f^{-}(\varepsilon) \, {\rm e}^{\beta \varepsilon}$ 
and
$
n^{+}(\varepsilon)=
n^{-}(\varepsilon) \, {\rm e}^{\beta \varepsilon}$, respectively. 

\subsection{Reformulation of the resonant tunneling approximation}
\label{sec:reformulation}

The reason why we adopt the generating functional approach 
is that once an approximate generating functional is obtained, 
one can calculate any order moment systematically 
by the functional derivation. 
In the following sections we perform practical calculations, 
employing $\bar{W}$ introduced in Sec. \ref{subsec:W}. 
Detailed discussions on our approximation are retained
in Appendix \ref{appendix:perturbartion}. 

The average current is calculated by
substituting $\bar{W}$ into Eq. (\ref{eqn:Ir})\cite{Utsumi_3}: 
\begin{eqnarray}
&I_{\rm r}&(t)
=
\frac{e}{\hbar}
\left. 
\frac{\delta \bar{W}}{\delta \varphi_{{\rm r} \, \Delta}(t)} 
\right|
\nonumber
\\
&=&
-e
\left.
{\rm Tr}
\left[
G_c
\frac{\delta \varphi_{\rm r}}{\delta \varphi_{{\rm r} \, \Delta}(t)}
\Sigma_{\rm r}
-
G_c
\Sigma_{\rm r}
\frac{\delta \varphi_{\rm r}}{\delta \varphi_{{\rm r} \, \Delta}(t)}
\right]
\right|
\label{eqn:trace1}.
\end{eqnarray}
Here, we used the fact that the self-energy Eq. (\ref{eqn:partselfenergy}) 
includes the phase factor through the 
particle-hole GF Eq. (\ref{eqn:particlehole}). 
In the language of Feynmann diagrams, Eq. (\ref{eqn:trace1}) can be rewritten 
in a compact form: 
\begin{eqnarray}
I_{\rm r}(t)
=-e
\left.
\left[
\begin{array}{c}
\includegraphics[width=7ex]{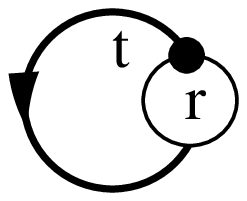}
\end{array}
-
\begin{array}{c}
\includegraphics[width=7ex]{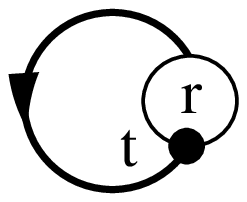}
\end{array}
\right]
\right|
\label{eqn:I1}.
\end{eqnarray}
Solid dots with $t$ represents 
$\delta \varphi_{\rm r}/ \delta \varphi_{{\rm r} \, \Delta}(t)$. 
The circle with ${\rm r}$ is the partial self-energy defined by
Eq. (\ref{eqn:partselfenergy}). 
Here we obtain the diagrammatic rule similar to that of Refs. 
\cite{Hershfield1,Hershfield2}.
\begin{description}
\item[(i)] \noindent
The diagrams corresponding to the functional derivative with respect 
to $\varphi_{{\rm r} \, \Delta}(t)$ is obtained by a series of 
the following operations;
to put a solid dot 
onto all possible positions of vertices in 
a closed diagram, 
to assign ${\rm r}$ on a circle connected to the solid dot, 
to multiple $\ri$, 
and 
to assign minus sign if a solid line comes into the solid dot. 
\end{description}

Next we project the fictitious time on $C$ to the real axis. 
As the tunneling Hamiltonian is zero on $C_{\tau}$, 
$\Sigma_{\rm r}(t,t')$ is zero for $t \in C_{\tau}$ or $t' \in C_{\tau}$. 
Hence, the diagrams in Eq. (\ref{eqn:I1}) are rewritten as, 
\begin{eqnarray}
{\rm Tr}
\left.
\left[
\tilde{G}_c
\frac{\delta \tilde{\varphi}_{\rm r}}
{\delta \varphi_{{\rm r} \, \Delta}(t)}
\tilde{\Sigma}_{\rm r}
\mtau^1
-
\tilde{G}
\mtau^1
\tilde{\Sigma}_{\rm r}
\frac{\delta \tilde{\varphi}_{\rm r}}
{\delta \varphi_{{\rm r} \, \Delta}(t)}
\right]
\right|,
\label{eqn:trace2}
\end{eqnarray}
where we performed the Keldysh rotation. 
$\mtau^{s} \: (s=0,1,2,3)$ is the Pauli matrix in the Keldysh space
\cite{Rammer}. 
The trace is carried out over the Keldysh space and products 
represents the integration along the real-time as
\begin{equation}
{\rm Tr} \left[ \tilde{g}_{\phi} \tilde{\varphi} \tilde{g}_{\phi} \right]
=
\int_{-\infty}^{\infty}
\rd 1 \rd 2 \,
{\rm Tr} \left[
\tilde{g}_{\phi}(1,2) \tilde{\varphi}(2) \tilde{g}_{\phi}(2,1)
\right].
\nonumber
\end{equation}
The phase in the physical representation 
$\tilde{\varphi}_{\rm r}$ is written as
$
\tilde{\varphi}_{\rm r}(t)
=
\varphi_{{\rm r} \, c}(t)\mtau^1+\varphi_{{\rm r} \, \Delta}(t)\mtau^0/2, 
$
which leads to a useful relation for the functional derivative technique:
\begin{equation}
\delta \tilde{\varphi}_{\rm r}(t')/\delta \varphi_{{\rm r} \, \Delta}(t)
=
\mtau^0\delta(t-t')/2.
\label{eqn:deltaphysical1r}
\end{equation}
By using the property of the GF in the physical representation 
$\tilde{G}(t,t')^{\dagger}=-\mtau^3 \tilde{G}(t',t) \mtau^3$, 
and that of a Pauli matrix $\mtau^3 \mtau^1 \mtau^3=-\mtau^1$, 
we can see that the second term of Eq. (\ref{eqn:trace2}) is 
minus the complex conjugate of the first term. 
To generalize this property, 
we obtain a helpful rule to reduce the number of diagrams: 
A diagram which is complex conjugate of a certain diagram 
is obtained by changing the direction of all lines 
and 
putting minus sign if the diagram includes odd numbers of 
vertices without solid dot. 

By performing the Fourier transformation, we rewrite 
Eq. (\ref{eqn:I1}) as:
\begin{eqnarray}
& &
2 e{\rm Re}
\left.
\begin{array}{c}
\includegraphics[width=7ex]{diagram11a.eps}
\end{array}
\right|
=
2 e{\rm Re}
\int \frac{\rd \varepsilon}{h}
{\rm Tr} \left[
\tilde{G}_c(\varepsilon) \frac{\mtau^0}{2} 
\tilde{\Sigma}_{\rm r}(\varepsilon) \mtau^1
\right]
\nonumber \\
\label{eqn:diagram-trace}
\\
&=&
e
\int
\frac{\rd \varepsilon}{h}
\frac{
\Sigma_{\rm r}^C(\varepsilon) G_c^K(\varepsilon)
}{2}
-
(C \leftrightarrow K).
\label{eqn:Iphysical}
\end{eqnarray}
The second term $(C \leftrightarrow K)$ is 
obtained from the first term by swapping superscripts $K$ and $C$. 
By generalizing  Eq. (\ref{eqn:diagram-trace}), 
we obtain the rule to calculate the diagram: 
\begin{description}
\item[(ii)] \noindent
Put $2 \times 2$ matrix GFs in the physical representation 
to the corresponding lines and circles, 
put $\mtau^0/2$ or $\mtau^1$ to a vertex with or without solid dot.
Subsequently, carry out the trace over Keldysh space and 
the integration over the frequency $\varepsilon/h$. 
\end{description}

Employing Eqs. (\ref{eqn:anti-comute}) and (\ref{eqn:comute}), 
Eq. (\ref{eqn:Iphysical}) is transformed 
into the single time representation:
$
(e/h)
\int \rd \varepsilon
\Sigma_{\rm r}^{+-}(\varepsilon) G_c^{-+}(\varepsilon)
-
(+ \leftrightarrow -).
$
By using Eqs. (\ref{eqn:correlationalpha}), (\ref{eqn:correlationself}) 
and (\ref{eqn:correlatonfullgc1}), 
and noting 
$I(t)=I_{\rm L}(t)=-I_{\rm R}(t)$,
we obtain the final form of the current expression 
which has the same form as the Landauer formula:
\begin{equation}
I(t)
=
\frac{1}{e R_{\rm K}}
\int \rd \varepsilon
T^F(\varepsilon)
\left \{
f^{-}_{\rm L}(\varepsilon)
-
f^{-}_{\rm R}(\varepsilon)
\right \},
\label{eqn:IRTA}
\end{equation}
where, $T^F$ is the effective transmission probability of lead electrons 
thorough the island:
\begin{eqnarray}
T^F(\varepsilon)
&=&
-
\alpha_{\rm L}^K(\varepsilon) \alpha_{\rm R}^K(\varepsilon)
|G_c^R(\varepsilon)|^2
\label{eqn:transmission}
\\
&=&
-
\frac{\alpha_{\rm L}^K(\varepsilon) \alpha_{\rm R}^K(\varepsilon)}
{\alpha^K(\varepsilon)}G_c^C(\varepsilon).
\label{eqn:transmission_2}
\end{eqnarray}
From the second form, we can see our result 
is equivalent to that of RTA\cite{Schoeller_Schon,note5}. 

In the same way as the average current, the average charge is 
evaluated by substitute $\bar{W}$ into Eq. (\ref{eqn:Q-exact})
\cite{Utsumi_4}, 
\begin{eqnarray}
\frac{Q(t)}{e}
&=&
\frac{1}{2}
-
\ri \hbar
\left.
\begin{array}{c}
\includegraphics[width=6ex]{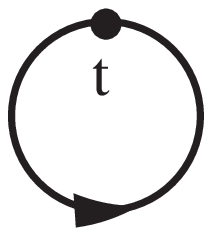}
\end{array}
\right|,
\nonumber
\end{eqnarray}
where the solid dot with $t$ corresponds to 
$\delta h/\delta h_{\Delta}(t)$ whose 
practical expression is equal to the right-hand side of 
Eq. (\ref{eqn:deltaphysical1r})
in the physical representation. 
The diagrammatic rule of functional derivation 
in terms of the scalar potential is given as follows:
\begin{description}
\item[(i')] \noindent
The diagrams corresponding to the functional derivative
with respect to $h(t)$ is obtained by 
inserting a solid dot into
all possible positions of $c$-field GF. 
\end{description}
Following the rule {\bf(ii)}, the diagram for average charge can be also 
calculated:
\begin{eqnarray}
\frac{Q(t)}{e}
&=&
\frac{1}{2}
+
\int \frac{\rd \varepsilon}{2 \ri \pi}
{\rm Tr} \left[ \tilde{G}_c(\varepsilon) \frac{\mtau^0}{2} \right]
\nonumber \\
&=&
\frac{1}{2}+\int \frac{\rd \varepsilon}{\pi}
\frac{\alpha^R(\varepsilon)}{\alpha^K(\varepsilon)}
{\rm Im} G_c^R (\varepsilon)
\label{eqn:QRTA}.
\end{eqnarray}

The imaginary part of $G_c^R$, 
which has a peak at $\varepsilon \sim \Delta_0$, 
describes the excitation property of the charge state. 
When the broadening of the peak is sufficiently small
and $\varepsilon,T,eV \ll E_C$, 
$G_c^R$ is approximately given by
\begin{equation}
G_c^R(\varepsilon)
\sim
z/
(\varepsilon-z \Delta_0+\ri \, z \, {\rm Im} \, \Sigma^R_c(z \, \Delta_0)).
\label{eqn:approxG}
\end{equation}
Here $z$ is the renormalization factor
$1/(1-
\left. 
\partial_\varepsilon {\rm Re} \Sigma^R_c(\varepsilon)
\right|_{\varepsilon=z \Delta_0})
\sim 1/(1+2 \alpha_0 \ln(E_C/\epsilon_C))$. 
The low energy cut-off $\epsilon_C$ is 
${\rm max}(|z \Delta_0|,2 \pi T, |eV|/2)$
where one of three parameters must be much larger 
than the other two\cite{Schoeller_text}. 
The imaginary part of self-energy ${\rm Im}\Sigma_c^R$ 
represents the life-time broadening effect. 
When the renormalization effect is negligible $z \sim 1$, 
it is written  as
$
{\rm Im} \, \Sigma^R_c(z \, \Delta_0)
\sim
\gamma
=
\hbar \Gamma/2
$,
where 
$\Gamma=\sum_{\rm r=L,R} (\Gamma_{\rm I \, r}+\Gamma_{\rm r \, I})$ 
and 
\begin{equation}
\Gamma_{\rm r \, I}
=
\frac{\rho(\Delta_0-\mu_{\rm r})}{e^2 R_{\rm r}}
 \; n^{-}(\Delta_0-\mu_{\rm r}),
\, \, 
\Gamma_{\rm I \, r}
=
\Gamma_{\rm r \, I}
{\rm e}^{-\beta \, (\Delta_0-\mu_{\rm r})}.
\label{eqn:tunnelingrate}
\end{equation}
$\Gamma_{\rm r \, I}$ ($\Gamma_{\rm I \, r}$) is equal to 
the tunneling rate into (out of) the island through the junction r
estimated by Fermi's golden rule. 
It is noticed that the condition $z \alpha_0 \ll 1$ 
is enough to neglect the broadening of the peak. 
By using the approximation 
Eq. (\ref{eqn:approxG}), Eq. (\ref{eqn:QRTA}) for equilibrium state 
reproduces the result of RTA\cite{Schoeller_Schon}, 
$
Q/e \sim
\left(
1
-z \tanh \left( z \Delta_0/(2 T) \right)
\right)/2$.

According to Ref. \cite{Schoeller_Schon}, 
Eq. (\ref{eqn:IRTA}) is consistent with the co-tunneling theory 
and orthodox theory for two-state limit\cite{Schon_R}. 
We can also confirm that Eq. (\ref{eqn:QRTA}) is
consistent with the orthodox theory in the following way: 
In the limit of $\alpha_0 \rightarrow 0$ 
with keeping $eV$ or $T$ finite, 
the renormalization factor approaches unity
and the imaginary part of Eq. (\ref{eqn:approxG}) reduces to 
the $\delta$-function $(-1/\pi) \, \delta(\varepsilon-\Delta_0)$. 
Thus, Eq. (\ref{eqn:QRTA}) reproduces the result of
the orthodox theory
\begin{equation}
\lim_{\alpha_0 \rightarrow 0}
Q(t)/e
=
\Gamma_{+}/\Gamma,
\end{equation}
where $\Gamma_{+/-}=\sum_{\rm r=L,R} \Gamma_{\rm r \, I / I \, r}$. 

\subsection{The current noise and the charge noise based on the 
reformulated resonant tunneling approximation 
}
\label{sec:mainbody}

Here, we calculate the current noise and the charge noise
based on the reformulated RTA. 
In the following discussions we limit ourselves to zero 
frequency component
\begin{equation}
S_{I{\rm r} \, I{\rm r'}}(0)=\int \rd t S_{I{\rm r} \, I{\rm r'}}(t,t'),
\end{equation}
where we used a fact that 
at a stationary state, the time translational invariance 
is satisfied and the correlation function depends only on 
the difference of $t$ and $t'$. 
From the definition Eq. (\ref{eqn:SII}), one obtain the 
current noise
by applying the rule {\bf(i)} twice. 
Reducing the number of diagrams and using the rule {\bf(ii)}, 
which can be also applied to the 
calculation of zero frequency noise diagram 
(Appendix. \ref{appendix:noiserule}), 
we obtain
\begin{widetext}
\begin{eqnarray}
& &
\frac{
S_{I{\rm r} \, I{\rm r'}}(0)
}{2 e^2}
=
\int \rd t
2
{\rm Re}
\left.
\left[
\begin{array}{c}
\includegraphics[width=8ex]{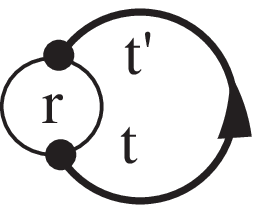}
\end{array}
\delta_{\rm r,r'}
+
\begin{array}{c}
\includegraphics[width=8ex]{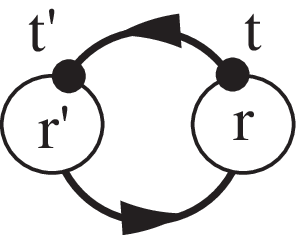}
\end{array}
-
\begin{array}{c}
\includegraphics[width=8ex]{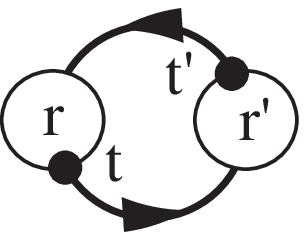}
\end{array}
\right]
\right|
-
\frac{(\Delta \rightarrow c)}{4}
\nonumber \\
&=&
2
\int \frac{\rd \varepsilon}{h}
{\rm Tr}
\left[
\frac{\mtau^0}{2} \tilde{\Sigma}_{\rm r}(\varepsilon) 
\frac{\mtau^0}{2} \tilde{G}_c(\varepsilon) 
\, \delta_{\rm r,r'}
+
\frac{\mtau^0}{2}
\tilde{\Sigma}_{\rm r}(\varepsilon) \mtau^1 
\tilde{G}_c(\varepsilon) \mtau^1 \tilde{\Sigma}_{\rm r'}(\varepsilon)
\frac{\mtau^0}{2}
\tilde{G}_c(\varepsilon)
-
\frac{\mtau^0}{2}
\tilde{\Sigma}_{\rm r}(\varepsilon) \mtau^1 \tilde{G}_c(\varepsilon)
\frac{\mtau^0}{2}
\tilde{\Sigma}_{\rm r'}(\varepsilon) \mtau^1 \tilde{G}_c(\varepsilon)
\right]
\nonumber \\
&-&
(\mtau^0 \rightarrow \mtau^1), 
\nonumber
\end{eqnarray}
where we used a fact that the second term of Eq. (\ref{eqn:SII}) 
is obtained from the first term by changing $\mtau^0$ to $\mtau^1$ 
in the physical representation 
(Appendix. \ref{appendix:noiserule}). 
After some straightforward calculations, 
we obtain the following expression:
\begin{eqnarray}
& &S_{I I}(0)
=
S_{I{\rm r} \, I{\rm \bar{r}}}(0)
=
-S_{I{\rm r} \, I{\rm r}}(0)
\nonumber 
\\
&=&
\frac{2}{R_{\rm K}}
\int \rd \varepsilon
\left[
-
\frac{\alpha_{\rm L}^K(\varepsilon) \alpha_{\rm R}^K(\varepsilon)}
{\alpha^K(\varepsilon)}G_c^C(\varepsilon)
\{
f^{-}_{\rm L}(\varepsilon) f^{+}_{\rm R}(\varepsilon)
+
f^{+}_{\rm L}(\varepsilon) f^{-}_{\rm R}(\varepsilon)
\}
-
\left \{
\frac{\alpha_{\rm L}^K(\varepsilon) \alpha_{\rm R}^K(\varepsilon)}
{\alpha^K(\varepsilon)}G_c^C(\varepsilon)
\right \}^2
\{
f^{-}_{\rm L}(\varepsilon)
-
f^{-}_{\rm R}(\varepsilon)
\}^2
\right]
\label{eqn:SIIRTA}
\\
&=&
\frac{2}{R_{\rm K}}
\int \rd \varepsilon
[
T^F(\varepsilon)
\{
f^{-}_{\rm L}(\varepsilon) f^{+}_{\rm R}(\varepsilon)
+
f^{+}_{\rm L}(\varepsilon) f^{-}_{\rm R}(\varepsilon)
\}
-
T^F(\varepsilon)^2
\{
f^{-}_{\rm L}(\varepsilon)
-
f^{-}_{\rm R}(\varepsilon)
\}^2
]
\label{eqn:SIIRTA_2},
\end{eqnarray}
where ${\rm \bar{r}}$ is the other side of ${\rm r}$
(for example, when the index $\rm r$ is $\rm L$ 
the index of the other side $\rm \bar{r}$ is $\rm R$). 
The charge noise is evaluated by adopting 
the definition Eq. (\ref{eqn:SQQ})
and by applying rules {\bf(i')} and {\bf(ii)}:
\begin{eqnarray}
& &
S_{Q Q}(0)
=
e^2 \hbar^2
\int \rd t
\left.
\begin{array}{c}
\includegraphics[width=7ex]{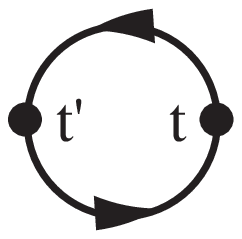}
\end{array}
\right|
-\frac{(\Delta \leftrightarrow c)}{4}
=
e^2 \hbar^2
\int \frac{\rd \varepsilon}{h}
{\rm Tr}
\left[
\tilde{G}_c(\varepsilon) \frac{\mtau^0}{2} 
\tilde{G}_c(\varepsilon) \frac{\mtau^0}{2}
\right]
-
(\mtau^0 \rightarrow \mtau^1)
\nonumber \\
&=&
e^2 \hbar^2
\int \frac{\rd \varepsilon}{h}
G_c^{-+}(\varepsilon)
G_c^{+-}(\varepsilon)
=
\sum_{\rm r,r'=L,R}
\frac{e^4 R_{\rm K}}{2 \pi^2}
\int \rd \varepsilon
\left|
G_c^R(\varepsilon)
\right|^4
\alpha^K_{\rm r}(\varepsilon)
\alpha^K_{\rm r'}(\varepsilon)
f^{+}_{\rm r}(\varepsilon)
f^{-}_{\rm r'}(\varepsilon),
\label{eqn:SQQRTA}
\end{eqnarray}
\end{widetext}
where we used Eqs. 
(\ref{eqn:correlationalpha}) 
and (\ref{eqn:correlatonfullgc1}). 

Equation (\ref{eqn:SIIRTA_2}) has the same form as the current 
noise expression of
a point contact without Coulomb interaction\cite{Lesovik}. 
This result is anticipated, because the tunneling current 
is expressed in the same form as the Landauer formula. 
However there is an important difference as mentioned in Ref. 
\cite{Schoeller_text}: 
The effective transmission probability includes
the Coulomb correlation and the inelastic relaxation effect. 

Here we note the following points.
First, our approximation satisfies the charge conservation law: 
As the approximate generating functional 
$\bar{W}$ is invariant under the gauge transformation 
Eq. (\ref{eqn:gauge_t}), 
we can show the conservation law, 
Eqs. (\ref{eqn:Iconserve}) and (\ref{eqn:Sconserve}), 
for the approximate expressions.
Especially for zero frequency component, 
they reduce to equations for the current conservation law, 
$\sum_{\rm r=L,R} I_{\rm r}=0$
and 
$\sum_{\rm r,r'=L,R} S_{I{\rm r} \, I{\rm r'}}(0)=0$. 
It is worth noticing that the gauge invariance 
is automatically satisfied when
$\bar{W}$ consists of closed diagrams. 
Secondly, we can show that our result satisfies 
the fluctuation-dissipation theorem: 
At $V=0$, the current noise expression is reduced to 
the Johnson-Nyquist formula
$S_{I I}(0)
=
4 \, T \, \bar{G}, 
$
where $\bar{G}$ is the conductance expressed as
$
\bar{G} = \lim_{V \rightarrow 0} \partial I/\partial V
=
R_{\rm K}^{-1} \int \rd \varepsilon
T^F(\varepsilon)/\{ 4 T \cosh \left( \varepsilon/(2 T) \right)^2 \}$. 


We discuss the current noise expression 
in the regime where $z \alpha_0 \ll 1$ 
and the renormalization effect is negligible, $z \sim 1$.
Using Eqs. (\ref{eqn:transmission}) and (\ref{eqn:approxG}), 
the first term of Eq. (\ref{eqn:SIIRTA}) 
is approximately given by
\begin{equation}
2e
\left( \gamma^{+}_{\rm r}(\gamma)
+
\gamma^{-}_{\rm r}(\gamma) \right),
\label{eqn:SIIcotwidth}
\end{equation}
where 
\begin{eqnarray}
\gamma^{+}_{\rm r}(\gamma)
&=&
\frac{R_{\rm K}}
{4 \pi^2 e R_{\rm L} R_{\rm R}}
\int \rd \varepsilon
\frac{
\rho(\varepsilon-\mu_{\rm r})
n^{-}_{\rm r}(\varepsilon)
\rho(\varepsilon-\mu_{\rm \bar{r}})
n^{+}_{\rm \bar{r}}(\varepsilon)
}
{|\varepsilon+\ri \gamma -\Delta_0|^2},
\nonumber \\
\gamma^{-}_{\rm r}(\gamma)
&=&
\gamma^{+}_{\rm r}(\gamma)
{\rm e}^{-\beta \, eV}.
\label{eqn:cot}
\end{eqnarray}
For $\gamma=0$, 
$\gamma^{+/-}_{\rm r}(0)$ is the co-tunneling current 
from lead r/$\bar{\rm r}$ to lead $\bar{\rm r}$/r, 
and 
Eq. (\ref{eqn:SIIcotwidth}) reproduces the co-tunneling theory
\cite{Averin,Brink,Sukhorukov} in the two-state limit. 
Equation (\ref{eqn:SIIcotwidth}), 
which we call the {\it co-tunneling theory with 
life-time broadening}, is equivalent to the 
previously proposed equation in Ref. \cite{KorotkovR}
in the limit of $T \rightarrow 0$. 
The validity of this approximation is discussed in the next section
from the point of view of the numerical results. 

By taking the limit $\alpha_0 \rightarrow 0$, with paying attention to
the relation
$
\lim_{\gamma \rightarrow 0}
(2 \pi \gamma)
\{ {\rm Im} [ 1/(\varepsilon-\Delta_0+\ri \gamma) ] \}^2
=
\delta(\varepsilon-\Delta_0)
$,
we obtain the result of the orthodox theory in the two-state limit: 
\begin{equation}
\lim_{\alpha_0 \rightarrow 0}
\left[
\alpha_0^{-1}
S_{I I}(0)
\right]
=
\alpha_0^{-1}
2 e
\left(
I_{+}-
\frac{2 I_{-}^2}{e \Gamma}
\right)
\label{eqn:SIIorth},
\end{equation}
where 
$
I_{\pm}
=
e
(
\Gamma_{\rm L \, I} \Gamma_{\rm I \, R}
\pm
\Gamma_{\rm I \, L} \Gamma_{\rm R \, I}
)/
\Gamma
$.
$I_{-}$ is the average current of the orthodox theory. 
The second term is related to the part of Eq. (\ref{eqn:SIIRTA}) 
proportional to $(T^{F})^2$,
which represents the Coulomb correlation and reduces the current noise 
from the Poissonian value. 
It is noticed that even at small tunneling conductance, the second term 
is important in ST regime. 

In the same way as above discussions, 
for $z \alpha_0 \ll 1 $ and $z \sim 1$, 
we obtain the co-tunneling theory\cite{Averin}
with life-time broadening
from 
Eq. (\ref{eqn:SQQRTA}):
\begin{eqnarray}
S_{Q Q}(0)
&\sim&
\sum_{\rm r,r'=L,R}
\frac{e^4 R_{\rm K}}{2 \pi^2}
\int \rd \varepsilon
\frac{
\alpha^K_{\rm r}(\varepsilon)
\alpha^K_{\rm r'}(\varepsilon)
f^{+}_{\rm r}(\varepsilon)
f^{-}_{\rm r'}(\varepsilon)
}
{
\{ (\varepsilon-\Delta_0)^2+\gamma^2 \}^2
}
.
\nonumber \\
\label{eqn:SQQcotwidth}
\end{eqnarray}
%
In the limit of $\alpha_0 \rightarrow 0$, 
we can confirm that our result reproduces the orthodox theory\cite{note4}:
\begin{equation}
\lim_{\alpha_0 \rightarrow 0}
\left[
\alpha_0
S_{Q Q}(0) 
\right]
=
\alpha_0
4 e^2
\Gamma_{+} \Gamma_{-}/
\Gamma^3
\label{eqn:SQQorth}.
\end{equation}

\section{results and discussions}
\label{sec:results}

In this section we present results obtained by the numerical calculation
of Eqs. (\ref{eqn:SIIRTA}) and (\ref{eqn:SQQRTA})\cite{note11},
based on which we discuss the 
non-equilibrium fluctuation in Sec. \ref{sec:discussion1}
and that of thermal fluctuation in Sec. \ref{sec:discussion2}. 
We also discuss the effect of renormalization 
on the noise
at low temperature caused by quantum fluctuation 
in Sec. \ref{sec:discussion3}. 

\subsection{Noise in the non-equilibrium state}
\label{sec:discussion1}

\begin{figure}[ht]
\includegraphics[width=.75 \columnwidth]{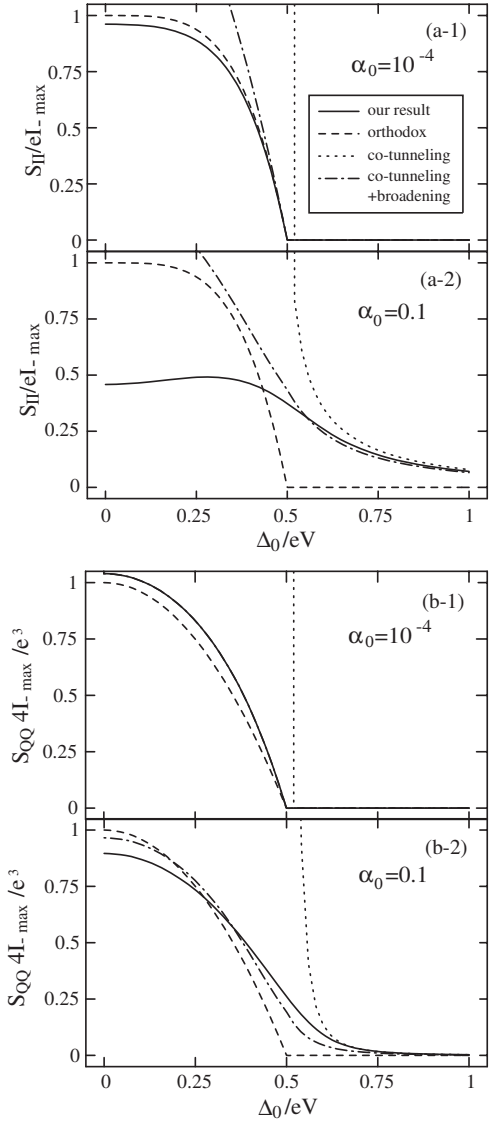}
\caption{
The excitation energy dependence of the current noise 
(the top panel)
and the charge noise (the bottom panel)
for $\alpha_0=10^{-4}$ ((a-1) and (b-1)) 
and $0.1$ ((a-2) and (b-2)) 
at 0K and $eV/E_C=0.4$. 
Plots are normalized by the value predicted by the orthodox theory 
at $\Delta_0=0$:
$e \, I_{- \, {\rm max}}=e \, G_0 \, V/2$ for the current noise
and 
$e^3/(4 I_{- \, {\rm max}})$ for the charge noise, 
where $G_0=1/(R_{\rm L}+R_{\rm R})$ is the series junction 
conductance. 
The solid, dashed, dotted and dot-dashed lines show the results 
evaluated by our approximation, 
orthodox theory, 
co-tunneling theory and
co-tunneling theory with life-time broadening, respectively.
In the panel (b-1), the solid line and dot-dashed line almost 
overlap each other. 
The parameters satisfy $eV \gg T_{\rm K}$; 
For example, $T_{\rm K}/E_C \sim 10^{-3}$ for $\alpha_0=0.1$. 
}
\label{fig:S}
\end{figure}

In order to clarify the nature of non-equilibrium current 
fluctuation, we consider the case of zero temperature, 
where there is no thermal fluctuation. 
Furthermore, we limit ourselves to the condition of the high bias voltage, 
$E_C \gg |eV| \gg T_{\rm K}$, where the renormalization effect is negligible. 
Figure \ref{fig:S} shows the current noise (the top panel)
and the charge noise (the bottom panel) as a function of the excitation energy
$\Delta_0$ for small $\alpha_0$ ((a-1) and (b-1)) and those for large 
$\alpha_0$ ((a-2) and (b-2)). 
Plots are normalized by values of the orthodox theory at $\Delta_0=0$. 
Solid lines, dashed lines and dotted lines show
results of our approximation
(Eqs. (\ref{eqn:SIIRTA}) and (\ref{eqn:SQQRTA})), 
the orthodox theory
(Eqs. (\ref{eqn:SIIorth}) and (\ref{eqn:SQQorth})) and
the co-tunneling theory
(Eqs. (\ref{eqn:SIIcotwidth}) and (\ref{eqn:SQQcotwidth})
with $\gamma=0$), respectively.

\begin{figure}[ht]
\includegraphics[width=.8 \columnwidth]{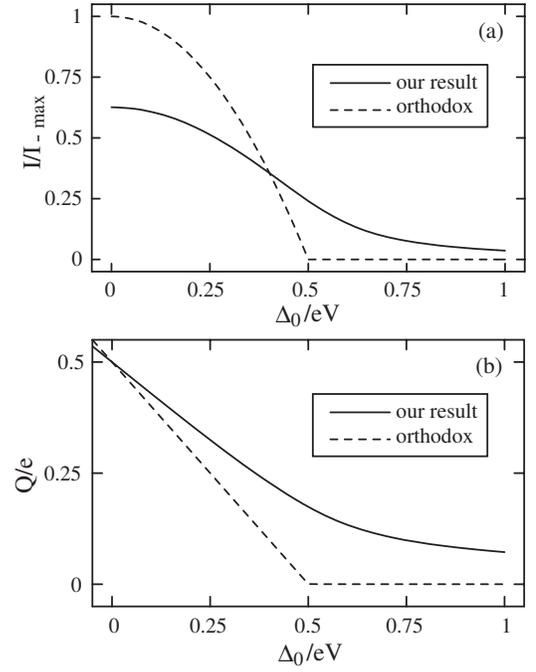}
\caption{
The excitation energy dependence of the 
normalized average current (a) and the 
average charge (b) for $\alpha_0=0.1$ at 0K and $eV/E_C=0.4$. 
The solid and dashed lines show the results of ours 
and those of orthodox theory, respectively. 
}
\label{fig:IN}
\end{figure}

When $\alpha_0$ is small, our results 
well reproduce the orthodox theory ((a-1) and (b-1)).
For large $\alpha_0$ ((a-2) and (b-2)) and in CB regime 
($|\Delta_0/(eV)| \gg 0.5$), 
they agree well with the co-tunneling theory. 
Figure \ref{fig:IN} shows the average current (a) and the average charge (b)
estimated by our approximation (solid lines) and 
the orthodox theory (dashed lines). 
In CB regime, we can see both the average value and the noise are
enhanced by the quantum fluctuation.
Around $\Delta_0=0$, 
the average current and the current noise are 
strongly suppressed due to the life-time broadening effect
\cite{Schoeller_Schon}.

Next, we discuss the validity of the co-tunneling theory with life-time 
broadening
(see Eqs. (\ref{eqn:SIIcotwidth}) and (\ref{eqn:SQQcotwidth});
dot-dashed lines in Fig. \ref{fig:S}). 
As for the charge noise in the limit of $\alpha_0 \rightarrow 0$, 
it reproduces the result of the orthodox theory as well as our approximation 
(Fig. \ref{fig:S} (b-1): 
The solid line and the dot-dashed line almost overlap each other). 
In ST regime it overestimates the current noise 
(Fig. \ref{fig:S} (a-1)), because it 
does not take the Coulomb correlation effect into account 
as mentioned before. 
We want to stress again that our result reproduces 
the orthodox theory in the limit of $\alpha_0 \rightarrow 0$, 
which can not be achieved by the co-tunneling theory with life-time 
broadening.

The physical picture of the non-equilibrium current fluctuation is
understood more clearly with the help of the Fano factor 
defined by $S_{II}/(2eI)$. 
The Fano factor is unity when the tunneling event of electrons is
Poissonian process and is suppressed when tunneling events 
are correlated. 
The orthodox theory predicts the sub-Poissonian behavior 
in ST regime because of the Coulomb correlation: 
Suppose one electron tunnels into the island through one junction, 
the next tunneling event must be the out going process of 
another electron 
through the other junction. 
In CB regime, the co-tunneling theory predict $S_{II}/(2eI)=1$.
It means that co-tunneling events, viz. the simultaneous tunneling events
of two electrons through the two junctions, occur randomly.

\begin{figure}[ht]
\includegraphics[width=.9 \columnwidth]{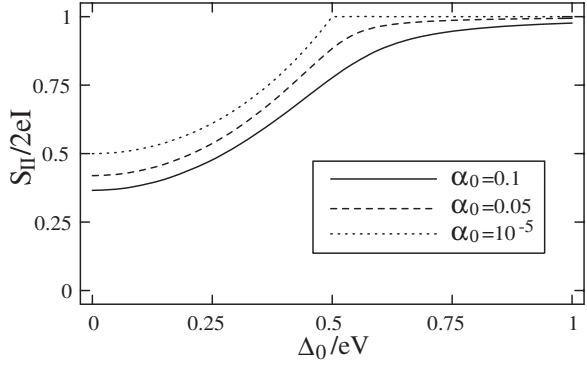}
\caption{
The excitation energy dependence of the Fano factor at $eV/E_C=0.4$ 
for $\alpha_0=10^{-5}$ (dotted line), 0.05 (dashed line) and 0.1 (solid line). 
}
\label{fig:fano}
\end{figure}

The figure \ref{fig:fano} shows the excitation energy dependence of 
the Fano factor obtained by our approximation.
In the small $\alpha_0$ limit, our approximation reproduces 
the orthodox theory in ST regime and the co-tunneling theory in CB regime
and smoothly interpolate two theories 
(The dotted line actually almost coincides 
with the result of orthodox theory in ST regime). 
For larger $\alpha_0$ the Fano factor is further suppressed
(the dashed and the solid line).
Especially, our result predicts the value smaller than $1/2$
at the degeneracy point. 
It is a distinctive result because the orthodox theory 
predicts the inequality $S_{II}/(2eI) \geq 1/2$
(p. 137 in Ref. \cite{Blanter}). 

Next we consider the physical meaning of our result. 
In CB regime and near the threshold voltage, 
the origin of the suppression 
of Fano factor is considered to be the enhancement of 
the effective transmission probability $T^F$, 
because the increase in the current noise is 
much larger then that in the charge noise
(see Figs. \ref{fig:S} (a-2) and \ref{fig:S} (b-2)). 
As the Fano factor of the shot noise is approximately given by
$1-T^F$\cite{Blanter}, the enhancement of the transmission probability 
results in the suppression of the Fano factor. 

Around the degeneracy point, the origin of the 
suppression of Fano factor is different, because the normalized current 
and the current noise are suppressed 
as shown in Fig. \ref{fig:IN} (a) and Fig. \ref{fig:S} (a-2). 
We consider that the Fano factor is suppressed because of
the dissipation, i.e. the life-time broadening effect: 
RTA takes account of the dissipation process 
which is the leak of an electron from the island while another electron 
tunnels into the island and relaxes to the local equilibrium
state of the island.
The suppression of the Fano factor by the
dissipation was previously predicted for the 1D electron channel coupled
with a boson bath\cite{Shimizu,Ueda}. 

\subsection{Effect of thermal fluctuation}
\label{sec:discussion2}

\begin{figure}[ht]
\includegraphics[width=.8 \columnwidth]{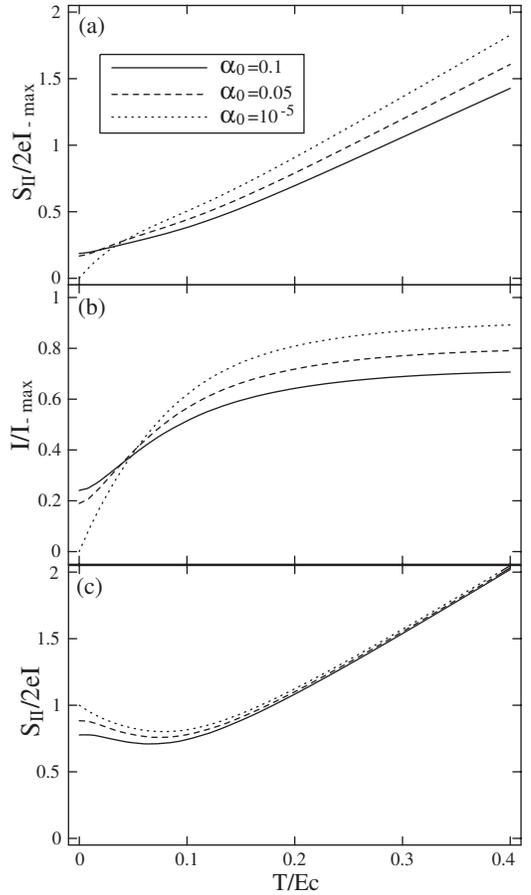}
\caption{
The temperature dependence of the normalized current noise (a) and 
the normalized average current (b) at a threshold $eV/2=\Delta_0=0.2 E_C$ 
for $\alpha_0=10^{-5}$ (dotted line), 0.05 (dashed line) and 0.1 (solid line). 
(c)
The temperature dependence of the Fano factor for various $\alpha_0$ at
the threshold. 
}
\label{fig:nyquist}
\end{figure}

Next we discuss the effect of the thermal fluctuation. 
Figure \ref{fig:nyquist} shows the temperature dependence of the current 
noise (a), the average current (b) and the Fano factor (c) at a 
threshold for various $\alpha_0$. 
They are normalized by the value of 
the orthodox theory at $T=\Delta_0=0$. 

As the temperature increases, the average current 
and the current noise increase because of the thermal fluctuation. 
At sufficiently high temperature $E_C \gg T \gg |eV|/2$, 
orthodox theory predicts that the average 
current saturates at $I_{- \, {\rm max}}$ 
and the thermal fluctuation dominates 
the current noise, i.e. $S_{II} \sim 4 (G_0/2) T$ 
which is the similar form as the Johnson-Nyquist noise 
for the ohmic resistance\cite{Zagoskin}
(the plot for $\alpha_0=10^{-5}$ almost coincides with the orthodox theory). 
Our result further shows that the average current and the current noise 
are suppressed as $\alpha_0$ increases (panels (a) and (b)). 
It is considered to be attributed to the higher order tunneling effect:
The lifetime broadening caused by the thermal fluctuation
is enhanced for the large tunnel conductance.
The panel (c) is the Fano factor versus temperature plot. 
The Fano factor is independent of $\alpha_0$,
which means that the correlation between tunneling events does not 
depend on $\alpha_0$.

\subsection{Renormalization effect}
\label{sec:discussion3}

\begin{figure}[ht]
\includegraphics[width=.8 \columnwidth]{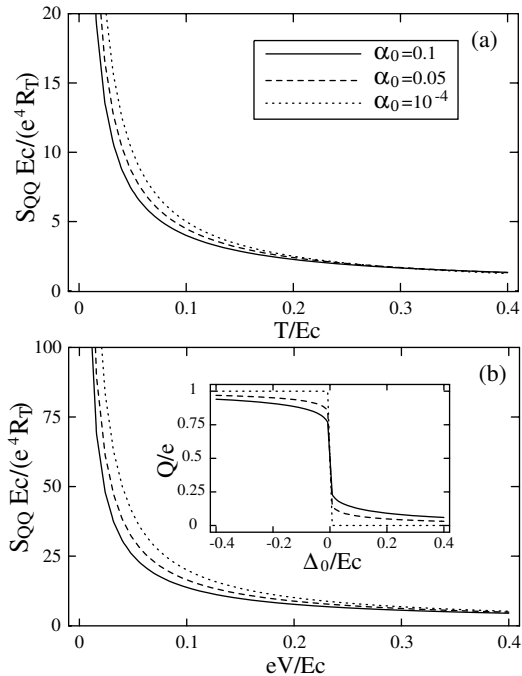}
\caption{
(a)The normalized charge noise as a function of temperature at 
$\Delta_0=eV=0$ for 
$\alpha_0=0.1$ (solid line), $0.05$ (dashed line) and $10^{-4}$ 
(dotted line). 
(b)The normalized charge noise as a function of bias voltage at 
$\Delta_0=T=0$.
Inset: The average charge as a function of $\Delta_0$ at $T=eV=0$. 
}
\label{fig:SQQ-low}
\end{figure}

Next we consider the renormalization effect at 
low bias voltage and temperature: $eV,T \lesssim T_{\rm K}$. 
Figures \ref{fig:SQQ-low} (a) and (b) show the charge 
noise normalized by $e^4 R_{\rm T}/E_C$ at $\Delta_0=0$ 
as a function of the temperature and the bias voltage, respectively. 
The charge noise is suppressed for large $\alpha_0$,
which is attributed to the renormalization of the 
system parameters.
In the regime $eV,T \ll T_{\rm K}$, where the life-time broadening effect 
is negligible ($z \alpha_0 \ll 1$), 
we can approximate Eq. (\ref{eqn:SQQRTA}) as 
\begin{equation}
4 (z \, e)^2 \tilde{\Gamma}_{+} \tilde{\Gamma}_{-}/ \tilde{\Gamma}^3,
\label{eqn:chargernormal}
\end{equation}
instead of Eq. (\ref{eqn:SQQorth}). 
Here $\tilde{\Gamma}$ and $\tilde{\Gamma}_{\pm}$ 
are tunneling rate Eq. (\ref{eqn:tunnelingrate}) 
written by using renormalized parameters such as
$\tilde{\Gamma}_{\rm r \, I}
=
z \, \rho(z \, \Delta_0-\mu_{\rm r})
\, n^{-}(z \, \Delta_0-\mu_{\rm r})/(e^2 R_{\rm r})$ where
the renormalization factor $z$ is
$1/(1+2 \alpha_0 \ln(E_C/\epsilon_C))$. 
The lower cut-off energy 
$\epsilon_C$ is $2 \pi T$ for the panel (a) and $|eV|/2$ for the panel (b). 
It is natural to interpret Eq. (\ref{eqn:chargernormal}) 
that the charge of a carrier is
modified as $z \, e$ by the renormalization effect. 
The interpretation is similar to that of 
the doubling of shot noise at 
the normal-metal(N)-superconductor(S) interface. 
Since at NS interface the carrier is
$2 e$-charged particle, viz. a Cooper pair,
the shot noise is twice as large as that at NN interface
\cite{Khlus}.

Though the normalized charge noise is suppressed 
with increasing $\alpha_0$, 
one see that the charge noise always diverges at 
$\Delta_0=T=eV=0$ for arbitrary $\alpha_0$ 
in the weak tunneling regime. 
Since the charge noise is related to the 
\lq \lq charge susceptibility" for excitation energy $\Delta_0$,
the divergence means that the number of charge 
changes by \lq \lq one" at the degeneracy point 
when we sweep the excitation energy. 
It is confirmed by the fact that the 
slope of excitation energy 
dependence of the average charge (an inset of Fig. \ref{fig:SQQ-low} (b))
diverges at the degeneracy point.

\begin{figure}[ht]
\includegraphics[width=.9 \columnwidth]{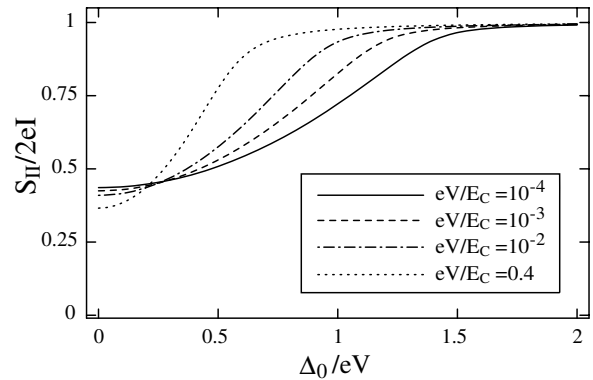}
\caption{The excitation energy dependence of the Fano factor 
for $\alpha_0=0.1$ at
$eV/E_C=10^{-4}$ (solid line), 
$10^{-3}$ (dashed line),
$10^{-2}$ (dot-dashed line) and 
$0.4$ (dotted line).}
\label{fig:fano3}
\end{figure}

Next we discuss the renormalization effect on the Fano factor. 
Figure \ref{fig:fano3} shows the excitation energy dependence 
of the Fano factor for various bias voltage. 
We can see at small bias voltage 
where the charging energy renormalization is
pronounced
$|eV| \lesssim T_{\rm K} \sim 10^{-3}$, 
the valley structures of curves are widened. 
The same behavior can be seen in the 
differential conductance shown in Refs. 
\cite{Schon_R,Konig_text,Schoeller_text}. 
We also see that the Fano factor is suppressed with
increasing bias voltage at $\Delta_0=0$. 
This suppression is also attributed to 
the dissipation as discussed in Sec. \ref{sec:discussion1}:
As the bias voltage increases, the dissipative charge fluctuation 
is enhanced and thus the Fano factor is suppressed. 

\section{summary}
\label{sec:summary}

%
%
By using the drone-fermion representation and the Schwinger-Keldysh approach, 
we have calculated the current noise and the charge noise 
in the regime of large quantum fluctuations of charge out of equilibrium. 
We have reformulated and extended RTA in a charge conserving way. 
Our approximation interpolates previous theories, 
the orthodox theory and the 
co-tunneling theory: 
Our result coincides with the orthodox theory in the limit of 
$\alpha_0 \rightarrow 0$ and 
is consistent with the co-tunneling theory in CB regime. 
The approximation is verified from the fact that 
the result satisfies the fluctuation-dissipation theorem. 
In previous papers, we also checked numerically 
that the energy sensitivity does not
exceed the quantum limit\cite{Utsumi_5,Utsumi_6}. 

%
%
We showed that 
at zero temperature and $E_C \gg |eV| \gg T_{\rm K}$,
the life-time broadening caused by non-equilibrium dissipative
charge fluctuation suppresses the current noise in ST regime. 
It also suppresses the Fano factor more than the Coulomb correlation does. 
Especially the Fano factor is suppressed below the minimum value predicted 
by the orthodox theory $1/2$ around $\Delta_0=0$. 
The origin of the suppression is attributed to 
the charge fluctuation which appears as 
the enhancement of the transmission probability in CB regime 
and the dissipation in ST regime.

%
%
At $E_C \gg T \gg |eV|/2 \gg T_{\rm K}$, 
we showed that the average current and 
the current noise deviate from the predictions of 
the orthodox theory with increasing $\alpha_0$. 
However, the Fano factor is independent of $\alpha_0$
and is proportional to the temperature. 
It means that the current noise is dominated by the thermal fluctuation 
and the correlation between the tunneling events does not depends on 
$\alpha_0$. 

%
%
At small bias voltage and temperature $eV,T \lesssim T_{\rm K}$, 
the charge noise is suppressed as compared with 
the prediction of the orthodox theory. 
We showed that it can be interpreted as the renormalization 
for the unit of island charge. 
Although the charge is renormalized, the 
charge noise diverges at $\Delta_0=T=eV=0$ for arbitrary $\alpha_0$
in the weak tunneling regime.
It means that the quantum fluctuation does not wash out the charge 
quantization. 

%
%
In this paper, we have limited ourselves to the discussions on the 
second moment and the zero frequency component, because 
we think them primitive. 
The investigation on the frequency dependence of noise will be important 
to estimate the performance of 
high-speed SET electrometer completely\cite{Johansson}. 
The investigation on the higher order moment and 
the full counting statistics\cite{Belzig1,Bagrets} will help us 
to understand carriers of strongly correlated system out 
of equilibrium.

\begin{acknowledgements}
We would like to thank Y. Isawa, J. Martinek, Yu. V. Nazarov 
and G. Johansson for variable discussions and comments. 
This work was supported by a Grant-in-Aid for Scientific
Research (C), No. 14540321 from MEXT. H.I. was supported by MEXT,
Grant-in-Aid for Encouragement of Young Scientists, No.
13740197.
\end{acknowledgements}

\appendix

\section{relation between generating functional 
representation
and operator representation}
\label{appendix:expressions}

In this Appendix, we show the relation between 
expressions for average and noise
in the operator representation and 
those in the generating functional representation. 
The variation of the exact action Eq. (\ref{eqn:action0}) 
accompanied by the infinitesimal variation of $h$ 
is given as the \lq \lq twisted" combination
(Sec. 9.3.2 in Ref. \cite{Chou}) of $Q$ and $h$, 
\begin{eqnarray}
\delta S
&=&
-
\int \rd t \, \frac{Q_c(t)}{e} \, \delta h_{\Delta}(t)
+
\left( c \leftrightarrow \Delta \right),
\label{eqn:dS}
\end{eqnarray}
where the center-of-mass coordinate of charge is
$Q_c(t)=e \, ( c_{+}(t)^*c_{+}(t)+c_{-}(t)^*c_{-}(t) )/2$. 
Employing this source term, we can show that the 
Eq. (\ref{eqn:Q-exact}) is equivalent to $\langle \hat{Q}(t) \rangle$
where $\hat{Q}=e \, (\hat{\sigma}_z+1)/2
=e \, \hat{c}^{\dag} \hat{c}$: 
\begin{eqnarray}
-e
\left.
\frac{\delta W}{\delta \, h_{\Delta}(t)}
\right|
&=&
\left.
\langle Q_c(t) \rangle
\right|
=
\langle \hat{Q}(t) \rangle
-
\frac{e}{2},
\nonumber
\end{eqnarray}
where we used Eq. (\ref{eqn:anti-comute}) to obtain the
final form. 

The second derivative 
$\left.
-\ri \hbar e^2 \delta^2 W
/
\delta h_{\Delta}(t) \delta h_{\Delta}(t')
\right|$ 
is calculated as 
\begin{equation}
\left.
\{
\langle Q_c(t) Q_c(t') \rangle
-
\langle Q_c(t) \rangle \langle Q_c(t') \rangle
\}
\right|
\label{eqn:SQQ_K}.
\end{equation}
Here the first term includes the correlation function 
of 4 field variables on a same branch, 
$\langle c_{\pm}^*(t) c_{\pm}(t) c_{\pm}^*(t') c_{\pm}(t') \rangle$, 
which is not well defined at $t=t'$. 
Usually, an additional operation to determine 
the order of field variables is required to remove the uncertainty. 
Alternatively, we subtract a term
\begin{equation}
\frac{1}{4}
\left.
\frac{-\ri \hbar e^2 \delta^2 W}
{\delta h_c(t) \delta h_c(t')}
\right|
=
\frac{
\left. 
\langle Q_{\Delta}(t) Q_{\Delta}(t') \rangle 
\right|}{4},
\label{eqn:SQQ2nd}
\end{equation}
from Eq. (\ref{eqn:SQQ_K}) as shown in our definition Eq. (\ref{eqn:SQQ}). 
Here we used the normalization Eq. (\ref{eqn:normalization}), 
$\left. \langle Q_{\Delta}(t) \rangle \right|=0$. 
As a result, the first term in  Eq. (\ref{eqn:SQQ_K})
is replaced by 
$(
\langle Q_{+}(t)Q_{-}(t')+Q_{-}(t)Q_{+}(t') \rangle
)/2
$,
which does not include the uncertainty. 
Using Eq. (\ref{eqn:anti-comute}), 
we show that our definition is equal to 
the standard charge noise expression
\begin{equation}
\left.
\frac{-\ri \hbar e^2 \delta^2 W}
{\delta h_{\Delta}(t) \delta h_{\Delta}(t')}
\right|
-\frac{(\Delta \rightarrow c)}{4}
=
\langle 
\{ \delta \hat{Q}(t), \delta \hat{Q}(t') \}
\rangle.
\nonumber
\end{equation}
We should stress that our definition does not change the final result
because Eq. (\ref{eqn:SQQ2nd}) is 0 from the normalization. 

The exact current expression is obtained in the same way. 
The source term corresponding to Eq. (\ref{eqn:dS}) is given by
$
\delta S
=
(\hbar/e)
\int
\rd t
\,
\sum_{\rm r=L,R}
I_{{\rm r} \, c}(t)
\,
\delta \varphi_{{\rm r} \, \Delta}(t)
+
\left(
c \leftrightarrow \Delta
\right), 
$
which leads to relations
\begin{equation}
I_{\rm r}(t)=\langle \hat{I}_{\rm r}(t) \rangle,
\, \,
S_{I{\rm r} \, I{\rm r'}}(t,t')
=
\langle \{ \delta \hat{I}_{\rm r}(t), \delta \hat{I}_{\rm r'}(t') \} \rangle,
\nonumber \\
\end{equation}
where the current operator at junction ${\rm r}$ is 
defined as
$
\hat{I}_{\rm r}(t)
=
(\ri e/\hbar)
\sum_{k n}
T_{\rm r}
{\rm e}^{\ri \varphi_{\rm r}(t)}
\hat{a}_{{\rm I} k n}^{\dag}
\hat{a}_{{\rm r} k n}
\hat{\sigma}_{+}
+{\rm h. c.}
$

\section{charge conservation}
\label{appendix:conservationlaw}

In this Appendix, we demonstrate
the charge conservation law. 
As $W$ is invariant under the transformation 
Eq. (\ref{eqn:gauge_t}), 
we obtain an identity
\begin{equation}
-e \partial_t \delta W/\delta \, h_{\Delta}(t)
=
(e/\hbar)
\sum_{\rm r=L,R}
\delta W/\delta \varphi_{{\rm r} \, \Delta}(t).
\label{eqn:identityW1}
\end{equation}
We can derive the other equation, which is 
obtained from above equation by replacing $\Delta$ with $c$. 
However, the latter equation is not important in 
the following discussions. 
By putting the auxiliary source fields as the values given in
the subscripts of Eq. (\ref{eqn:I-exact}), and 
employing
Eqs. (\ref{eqn:I-exact}) and (\ref{eqn:Q-exact}), 
we obtain the current continuity equation, 
Eq. (\ref{eqn:Iconserve}). 

Next we demonstrate the charge conservation law for correlation functions. 
By acting the operator 
$ \ri e \hbar \, \partial_{t'} \, \delta / \delta h_{\Delta}(t')$ 
or 
$ -\ri e \sum_{\rm r'=L,R} \delta/ \delta \varphi_{{\rm r} \Delta}(t')$ 
on Eq. (\ref{eqn:identityW1}), we obtain following two equations: 
\begin{eqnarray}
-\partial_{t'} \, \partial_t \, 
\frac{\ri \hbar \, e^2 \, \delta^2 W}
{\delta \, h_{\Delta}(t') \, \delta \, h_{\Delta}(t)}
&=&
\partial_{t'}
\sum_{\rm r=L,R}
\frac{\ri e^2 \, \delta^2 W}
{\delta \, h_{\Delta}(t') \, \delta \, \varphi_{{\rm r} \Delta}(t)},
\nonumber \\
\partial_t \sum_{\rm r'=L,R}
\frac{\ri e^2 \, \delta^2 W}
{\delta \varphi_{{\rm r'} \Delta}(t') \, \delta h_{\Delta}(t)}
&=&
\sum_{\rm r,r'=L,R}
\frac{e^2}{\ri \hbar}
\frac{\delta^2 W}
{\delta \varphi_{{\rm r'} \Delta}(t')
\,
\delta \varphi_{{\rm r} \Delta}(t)}. 
\nonumber
\end{eqnarray}
By comparing the left-hand side of the former equation and 
the right-hand side of the latter equation, 
by setting the auxiliary source fields as the values given in
the subscripts of Eq. (\ref{eqn:I-exact}), 
and by using
Eqs. (\ref{eqn:SII}) and (\ref{eqn:SQQ}), 
we obtain the charge conservation law for correlation 
functions, Eq. (\ref{eqn:Sconserve}).

\section{loop diagrams: particle-hole Green 
function and self-energy}
\label{appendix:loop}

In this Appendix, we calculate the particle-hole GF.
We begin with the tunneling action for the large transverse channel 
obtained from Eq. (\ref{eqn:action0})
by tracing out the electron degrees of freedom\cite{Utsumi_3}:
$S_{\rm T}
=
\int _C \rd 1 \rd 2
\, \sigma_{+}(1) \alpha(1,2) \sigma_{-}(2)$.
In the physical representation, it is rewritten as
$
\int \rd 1 \rd 2 \;
\tilde{\vec{\sigma}}_{+}(1)^{\dagger}
\mtau^1
\tilde{\alpha}_{\rm r}(1,2)
\mtau^1
\tilde{\vec{\sigma}}_{-}(2)
$,
where the vector field $\tilde{\vec{\sigma}}_{\pm}$ 
is defined in the same way as Eq. (\ref{eqn:fieldphysical}). 
Each component of $\tilde{\alpha}_{\rm r}$ can be 
calculated by utilizing the functional derivation. 
For example, (1,2)-component is 
\begin{eqnarray}
& &\left( \mtau^1 \tilde{\alpha}_{\rm r}(1,2) \mtau^1 \right)_{1,2}
=
\frac{\delta^2 S_{\rm T}}
{\delta \sigma_{+ \, 1}(1)^{*}
 \delta \sigma_{- \, 2}(2)} 
\nonumber \\
&=&
-\ri \hbar
N_{\rm ch} T_{\rm r}^2
\,
{\rm Tr}[
\frac{\delta \sigma_{+}^{*}}{\delta \sigma_{+ \, 1}(1)^{*}}
{\rm e}^{\ri \varphi_{\rm r}}
g_{\rm r}
\frac{\delta \sigma_{-}}{\delta \sigma_{- \, 2}(2)}
{\rm e}^{\ri \varphi_{\rm r}}
g_{\rm I}
]. 
\nonumber \\
\label{eqn:traceA1}
\end{eqnarray}
As the functional derivative is 
$\delta \tilde{\sigma}_{\pm}(t')/\delta \sigma_{{\pm} \, 1(2)}(t)
=\mtau^{0(1)} \delta(t-t')/\sqrt{2}$
in the physical representation, 
the trace yields
${\rm e}^{\ri \kappa_{\rm r} eV (t_2-t_1)/\hbar}
{\rm Tr}[\mtau^0 \tilde{g}_{\rm r}(1,2) \mtau^1 \tilde{g}_{\rm I}(2,1)]/2$. 
Here we put $\varphi_{\Delta}=0$ and $\varphi_c(t)=eVt/\hbar$. 
The other components are evaluated in the same way.
Then four components are given as
\begin{eqnarray}
& &
\begin{pmatrix}
0 &
g_{\rm r}^A g_{\rm I}^K+g_{\rm r}^K g_{\rm I}^R \\
g_{\rm r}^R g_{\rm I}^K+g_{\rm r}^K g_{\rm I}^A &
g_{\rm r}^K g_{\rm I}^K-g_{\rm r}^C g_{\rm I}^C  
\end{pmatrix}
\label{eqn:ph},
\end{eqnarray}
where we omitted arguments and coefficients. 
To obtain this form, we used the normalization 
Eq. (\ref{eqn:normalization}) or the relation
$\theta(t) \, \theta(-t)=0$ 
(see Eqs. (2.64) and (2.65) in Ref. \cite{Chou}). 
The following calculations are same as those in Ref. \cite{Utsumi_3}. 
Employing the Fourier transforms of the GF defined in
Eq. (\ref{eqn:GFl}), 
$g^R_{\rm r}(\varepsilon) = - \ri \pi \rho_{\rm r}(\varepsilon)$ 
and
$g^K_{\rm r}(\varepsilon) = -2 \ri \pi \rho_{\rm r}(\varepsilon)
\tanh ( \varepsilon/(2 T))$, 
we obtain Eq. (\ref{eqn:alphaRK}). 

Another loop diagram, the self-energy 
Eq. (\ref{eqn:partselfenergy}), can be calculated in the same way. 
Four components are given in the same form 
as those of Eq. (\ref{eqn:ph}). 
By using Eqs. (\ref{eqn:dRK}) and (\ref{eqn:alphaRK}), we obtain 
\begin{equation}
\Sigma_{\rm r}^R(\varepsilon)
=
\int
\frac{\rd \varepsilon'}{2 \pi}
\frac{\ri \, \alpha_{\rm r}^K(\varepsilon')}
{\varepsilon+\ri \eta -\varepsilon'}, 
\; \; 
\Sigma_{\rm r}^K(\varepsilon)
=
\alpha_{\rm r}^C(\varepsilon).
\nonumber
\end{equation}
From these equations, Eqs. (\ref{eqn:selfenergy}) and (\ref{eqn:selfenergyR})
can be derived.

\section{perturbation theory}
\label{appendix:perturbartion}

In this Appendix, we describe some results of finite order 
perturbation theory
and explain why we introduced Eq. (\ref{eqn:WRTA})
which takes account of all orders for $c$-field corrections. 
First, we calculate the first order contribution of 
average charge by employing $W^{(1)}$, Eq. (\ref{eqn:Q-exact}) and
rules {\bf(i')} and {\bf(ii)}.  
\begin{widetext}
\begin{eqnarray}
& &
\frac{
Q^{(1)}(t)
}{e}
=
-\ri \hbar
\left.
\begin{array}{c}
\includegraphics[width=8ex]{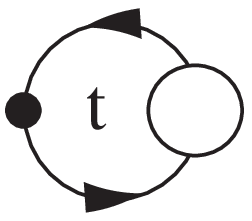}
\end{array}
\right|
=
-\frac{\ri}{2 \pi}
\int \rd \varepsilon
{\rm Tr} \left[ \frac{\mtau^0}{2}
\tilde{g}_c(\varepsilon) \mtau^1 
\tilde{\Sigma}_c(\varepsilon) \mtau^1 
\tilde{g}_c(\varepsilon) 
\right]
\nonumber \\
&=&
-\ri
\int
\frac{\rd \varepsilon}{8 \pi}
(
g_c^K(\varepsilon) g_c^P(\varepsilon) \Sigma_c^P(\varepsilon)
+g_c^K(\varepsilon) g_c^C(\varepsilon) \Sigma_c^C(\varepsilon)+
2 |g_c^R(\varepsilon)|^2 \Sigma_c^K(\varepsilon)
),
\nonumber
\end{eqnarray}
\end{widetext}
where GF denoted with superscript $P$ is given 
as $g_c^P=g_c^R+g_c^A$ etc. 
In the equilibrium state, the second and the third terms of
the second line are negligibly small $O(\Delta_0/E_C)$ 
and the first term is simplified to
\begin{equation}
\frac{
Q^{(1)}
}{e}
\sim
\frac{1}{2}
\partial_{\Delta_0}
\left \{
\tanh \left( \frac{\Delta_0}{2 T} \right)
{\rm Re}
\Sigma_c^R(\Delta_0)
\right \},
\label{eqn:Q2nd}
\end{equation}
where we utilized the relation 
$g_c^K(\varepsilon) g_c^P(\varepsilon)=
\partial_{\Delta_0} g_c^K(\varepsilon)$. 
In the limit of zero temperature, Eq. (\ref{eqn:Q2nd}) 
leads to 
the log-divergence\cite{Schon_R,Matveev_CB} as
$\sim -\alpha_0 \ln (E_C/|\Delta_0|) \, {\rm sgn}(\Delta_0)$. 

Above result suggests that the $c$-field correction 
is responsible for the divergence. 
It is further confirmed by calculating second order 
contribution of average 
current generated from $c$-field correction $W_{c{\rm -field}}^{(2)}$. 
Employing rules {\bf(i)} and {\bf(ii)}, we obtain
\begin{eqnarray}
& &I_{\rm r}^{(2)}(t)
=
2 e
{\rm Re}
\left.
\begin{array}{c}
\includegraphics[width=8ex]{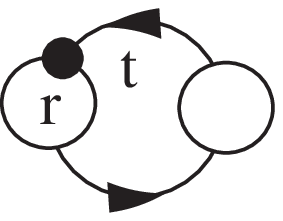}
\end{array}
\right|
\nonumber \\
&=&
2 e
{\rm Re}
\int \frac{\rd \varepsilon}{h}
{\rm Tr}
\left[
\frac{\mtau^0}{2}
\tilde{\Sigma}_{\rm r}(\varepsilon) 
\mtau^1
\tilde{g}_c(\varepsilon) \mtau^1 
\tilde{\Sigma}_c(\varepsilon) \mtau^1 
\tilde{g}_c(\varepsilon)
\right]
\nonumber \\
&=&
e
\int \frac{\rd \varepsilon}{2 h}
|g_c^R(\varepsilon)|^2
\Sigma^K_{\rm \bar{r}}(\varepsilon)
\Sigma^C_{\rm r}(\varepsilon)-(C \leftrightarrow K)
+
\delta I^{(2)}_{\rm r}
\nonumber \\
&=&
\gamma^{+}_{\rm r}(\eta)
-
\gamma^{-}_{\rm r}(\eta)
+
\delta I^{(2)}_{\rm r}. 
\label{eqn:I2}
\end{eqnarray}
The first and the second terms of Eq. (\ref{eqn:I2}), 
which are consistent of the expression 
for co-tunneling current \cite{Averin_Nazarov},
also diverge at the degeneracy point. 
From above discussions, we can deduce that the most simple way 
to regulate the divergence 
is to sum up $c$-field corrections $(g_c \Sigma_c)^n$ 
up to infinite $n$ as shown in Eq. (\ref{eqn:WRTA}). 
It should be noted that the correction term 
$\delta I_{\rm r}^{(2)}=
e
\int \rd \varepsilon
\,
g_c^C(\varepsilon) \Sigma_{\rm r}^C(\varepsilon)
\{
g_c^C(\varepsilon) \Sigma_{\rm r}^K(\varepsilon)
-g_c^K(\varepsilon) \Sigma_c^C(\varepsilon)
\}
/(4 h)
\sim
O(1/\eta)
$
diverges in the limit of $\eta \rightarrow 0$. 
This divergence disappears when we 
consider Eq. (\ref{eqn:WRTA})
as discussed in 
Sec. \ref{sec:reformulation} and Sec. \ref{sec:mainbody}.

\section{rule for calculation of zero frequency noise diagrams}
\label{appendix:noiserule}

In this Appendix, we demonstrate that the rule 
{\bf(ii)} can be also applied to the calculation 
of zero-frequency noise. 
We also demonstrate the rule to calculate the second 
term of the noise expression
Eq. (\ref{eqn:SII}) or Eq. (\ref{eqn:SQQ}). 
For example, we consider the current noise 
related to $W^{(1)}$.
Though we consider the simple case, the following discussions 
can be generalized. 
From the definition Eq. (\ref{eqn:SIIpart}), 
the noise diagrams is obtained as: 
\begin{equation}
\frac{
S^{(1)}_{I{\rm r} \, I{\rm r'}}(0)
}{2 e^2}
=
\int \rd t
2{\rm Re}
\left.
\begin{array}{c}
\includegraphics[width=7ex]{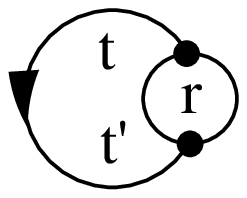}
\end{array}
\right|
\delta_{\rm r,r'}
-
\frac{(\Delta \rightarrow c)}{4}.
\label{eqn:SII_1}
\end{equation}
The integration in terms of $t$ in the first term is calculated as
\begin{eqnarray}
& &
\int \rd t
\left.
\begin{array}{c}
\includegraphics[width=7ex]{diagram17.eps}
\end{array}
\right|
=
\int \rd t
{\rm Tr}
\left.
\left[
\frac{\delta \varphi_{\rm r}}{\delta \varphi_{{\rm r} \, \Delta}(t)}
\Sigma_{\rm r}
\frac{\delta \varphi_{\rm r'}}{\delta \varphi_{{\rm r'} \, \Delta}(t')}
g_c
\right]
\right|
\nonumber \\
&=&
\int \frac{\rd \varepsilon}{h}
{\rm Tr}
\left[
\frac{\mtau^0}{2}
\tilde{\Sigma}_{\rm r}(\varepsilon)
\frac{\mtau^0}{2}
\tilde{g}_c(\varepsilon)
\right].
\label{eqn:trace4}
\end{eqnarray}
And thus we can see that the rule {\bf(ii)} can be applied
for the calculation of the zero-frequency current noise. 
As for the second term, the derivation
\begin{equation}
\delta \tilde{\varphi}_{\rm r}(t')/\delta \varphi_{{\rm r} \, c}(t)
=
\mtau^1\delta(t-t'),
\label{eqn:deltaphysical1c}
\end{equation}
appears instead of Eq. (\ref{eqn:deltaphysical1r}). 
Thus we can derive the rule that the second term
is obtained from the first term by replacing $\mtau^0$ with $\mtau^1$ as
$\int \rd \varepsilon
\,
2{\rm Re}
{\rm Tr}
\left[
(\mtau^1/2) 
\tilde{\Sigma}_{\rm r}(\varepsilon) 
(\mtau^1/2)
\tilde{g}_c(\varepsilon)
\right]/h
$
\cite{note3}. 
As a result, 
Eq. (\ref{eqn:SII_1}) is expressed as
\begin{equation}
{\rm Re} \int \rd \varepsilon
{\rm Tr} 
\left[ \mtau^0 
\tilde{\Sigma}_{\rm r}(\varepsilon) 
\mtau^0 \tilde{g}_c(\varepsilon) 
\right]
\delta_{\rm r,r'}/(2 h)
-
(\mtau^0 \rightarrow \mtau^1).
\nonumber
\end{equation}
As for the charge noise, we can repeat the same 
discussions as above.


\begin{thebibliography}{63}
\expandafter\ifx\csname natexlab\endcsname\relax\def\natexlab#1{#1}\fi
\expandafter\ifx\csname bibnamefont\endcsname\relax
  \def\bibnamefont#1{#1}\fi
\expandafter\ifx\csname bibfnamefont\endcsname\relax
  \def\bibfnamefont#1{#1}\fi
\expandafter\ifx\csname citenamefont\endcsname\relax
  \def\citenamefont#1{#1}\fi
\expandafter\ifx\csname url\endcsname\relax
  \def\url#1{\texttt{#1}}\fi
\expandafter\ifx\csname urlprefix\endcsname\relax\def\urlprefix{URL }\fi
\providecommand{\bibinfo}[2]{#2}
\providecommand{\eprint}[2][]{\url{#2}}

\bibitem[{\citenamefont{Averin and Likharev}(1991)}]{Averin_Likharev}
\bibinfo{author}{\bibfnamefont{D.~V.} \bibnamefont{Averin}} \bibnamefont{and}
  \bibinfo{author}{\bibfnamefont{K.~K.} \bibnamefont{Likharev}}, in
  \emph{\bibinfo{booktitle}{Mesoscopic Phenomena in Solids}}, edited by
  \bibinfo{editor}{\bibfnamefont{B.~L.} \bibnamefont{Altshuler}},
  \bibinfo{editor}{\bibfnamefont{P.~A.} \bibnamefont{Lee}}, \bibnamefont{and}
  \bibinfo{editor}{\bibfnamefont{R.~A.} \bibnamefont{Webb}}
  (\bibinfo{publisher}{Elsevier Science Publishers B. V.},
  \bibinfo{address}{Amsterdam}, \bibinfo{year}{1991}),
  vol.~\bibinfo{volume}{30} of \emph{\bibinfo{series}{Modern Problem in
  Condensed Matter Sciences}}, chap.~\bibinfo{chapter}{6}.

\bibitem[{\citenamefont{Ingold and Nazarov}(1992)}]{Ingold_Nazarov}
\bibinfo{author}{\bibfnamefont{G.-L.} \bibnamefont{Ingold}} \bibnamefont{and}
  \bibinfo{author}{\bibfnamefont{Y.~V.} \bibnamefont{Nazarov}}, in
  \cite{Sin-Cha-Tun}, chap.~\bibinfo{chapter}{2}, p.~\bibinfo{pages}{21}.

\bibitem[{\citenamefont{Sch{\"o}n}(1998)}]{Schon_R}
\bibinfo{author}{\bibfnamefont{G.}~\bibnamefont{Sch{\"o}n}}, in
  \emph{\bibinfo{booktitle}{Quantum Transport and Dissipation}}, edited by
  \bibinfo{editor}{\bibfnamefont{T.}~\bibnamefont{Dittrich}},
  \bibinfo{editor}{\bibfnamefont{P.}~\bibnamefont{H{\"a}nggi}},
  \bibinfo{editor}{\bibfnamefont{G.-L.} \bibnamefont{Ingold}},
  \bibinfo{editor}{\bibfnamefont{B.}~\bibnamefont{Kramer}},
  \bibinfo{editor}{\bibfnamefont{G.}~\bibnamefont{Sch{\"o}n}},
  \bibnamefont{and} \bibinfo{editor}{\bibfnamefont{W.}~\bibnamefont{Zwerger}}
  (\bibinfo{publisher}{Wiley-VCH}, \bibinfo{address}{Weinheim},
  \bibinfo{year}{1998}), p. \bibinfo{pages}{149}.

\bibitem[{\citenamefont{Matveev}(1991)}]{Matveev_CB}
\bibinfo{author}{\bibfnamefont{K.~A.} \bibnamefont{Matveev}},
  \bibinfo{journal}{Sov. Phys. JETP.} \textbf{\bibinfo{volume}{72}},
  \bibinfo{pages}{892} (\bibinfo{year}{1991}).

\bibitem[{\citenamefont{Falci et~al.}(1995)\citenamefont{Falci, Sch{\"o}n, and
  Zimanyi}}]{Falci_2}
\bibinfo{author}{\bibfnamefont{G.}~\bibnamefont{Falci}},
  \bibinfo{author}{\bibfnamefont{G.}~\bibnamefont{Sch{\"o}n}},
  \bibnamefont{and} \bibinfo{author}{\bibfnamefont{G.~T.}
  \bibnamefont{Zimanyi}}, \bibinfo{journal}{Phys. Rev. Lett.}
  \textbf{\bibinfo{volume}{74}}, \bibinfo{pages}{3257} (\bibinfo{year}{1995}).

\bibitem[{\citenamefont{Schoeller and Sch{\"o}n}(1994)}]{Schoeller_Schon}
\bibinfo{author}{\bibfnamefont{H.}~\bibnamefont{Schoeller}} \bibnamefont{and}
  \bibinfo{author}{\bibfnamefont{G.}~\bibnamefont{Sch{\"o}n}},
  \bibinfo{journal}{Phys. Rev. B} \textbf{\bibinfo{volume}{50}},
  \bibinfo{pages}{18436} (\bibinfo{year}{1994}).

\bibitem[{\citenamefont{Schoeller and K{\"o}nig}(2000)}]{Schoeller_Konig}
\bibinfo{author}{\bibfnamefont{H.}~\bibnamefont{Schoeller}} \bibnamefont{and}
  \bibinfo{author}{\bibfnamefont{J.}~\bibnamefont{K{\"o}nig}},
  \bibinfo{journal}{Phys. Rev. Lett.} \textbf{\bibinfo{volume}{84}},
  \bibinfo{pages}{3686} (\bibinfo{year}{2000}).

\bibitem[{\citenamefont{K{\"o}nig et~al.}(1995)\citenamefont{K{\"o}nig,
  Schoeller, Sch{\"o}n, and Fazio}}]{Konig_text}
\bibinfo{author}{\bibfnamefont{J.}~\bibnamefont{K{\"o}nig}},
  \bibinfo{author}{\bibfnamefont{H.}~\bibnamefont{Schoeller}},
  \bibinfo{author}{\bibfnamefont{G.}~\bibnamefont{Sch{\"o}n}},
  \bibnamefont{and} \bibinfo{author}{\bibfnamefont{R.}~\bibnamefont{Fazio}}, in
  \emph{\bibinfo{booktitle}{Quantum Dynamics of Submicron Structures}}, edited
  by \bibinfo{editor}{\bibfnamefont{H.~A.} \bibnamefont{Cerdeira}},
  \bibinfo{editor}{\bibfnamefont{B.}~\bibnamefont{Kramer}}, \bibnamefont{and}
  \bibinfo{editor}{\bibfnamefont{G.}~\bibnamefont{Sch{\"o}n}}
  (\bibinfo{publisher}{Kluwer Academic Publishers},
  \bibinfo{address}{Dordrecht}, \bibinfo{year}{1995}), vol.
  \bibinfo{volume}{291} of \emph{\bibinfo{series}{NATO ASI Series E}}, p.
  \bibinfo{pages}{221}, \bibinfo{note}{and references therein.}

\bibitem[{\citenamefont{Golubev and Zaikin}(1994)}]{Golubev_Zaikin}
\bibinfo{author}{\bibfnamefont{D.~S.} \bibnamefont{Golubev}} \bibnamefont{and}
  \bibinfo{author}{\bibfnamefont{A.~D.} \bibnamefont{Zaikin}},
  \bibinfo{journal}{Phys. Rev. B} \textbf{\bibinfo{volume}{50}},
  \bibinfo{pages}{8736} (\bibinfo{year}{1994}).

\bibitem[{\citenamefont{Joyez et~al.}(1997)\citenamefont{Joyez, Bouchiat,
  Esteve, Urbina, and Devoret}}]{Joyez}
\bibinfo{author}{\bibfnamefont{P.}~\bibnamefont{Joyez}},
  \bibinfo{author}{\bibfnamefont{V.}~\bibnamefont{Bouchiat}},
  \bibinfo{author}{\bibfnamefont{D.}~\bibnamefont{Esteve}},
  \bibinfo{author}{\bibfnamefont{C.}~\bibnamefont{Urbina}}, \bibnamefont{and}
  \bibinfo{author}{\bibfnamefont{M.~H.} \bibnamefont{Devoret}},
  \bibinfo{journal}{Phys. Rev. Lett.} \textbf{\bibinfo{volume}{79}},
  \bibinfo{pages}{1349} (\bibinfo{year}{1997}).

\bibitem[{\citenamefont{Schoeller}(1997)}]{Schoeller_text}
\bibinfo{author}{\bibfnamefont{H.}~\bibnamefont{Schoeller}}, in
  \emph{\bibinfo{booktitle}{Mesoscopic Electron Transport}}, edited by
  \bibinfo{editor}{\bibfnamefont{L.~L.} \bibnamefont{Sohn}},
  \bibinfo{editor}{\bibfnamefont{L.~P.} \bibnamefont{Kouwenhoven}},
  \bibnamefont{and} \bibinfo{editor}{\bibfnamefont{G.}~\bibnamefont{Sch{\"o}n}}
  (\bibinfo{publisher}{Kluwer Academic Publishers},
  \bibinfo{address}{Dordrecht}, \bibinfo{year}{1997}), vol.
  \bibinfo{volume}{345} of \emph{\bibinfo{series}{NATO ASI Series E}}, p.
  \bibinfo{pages}{291}.

\bibitem[{\citenamefont{Blanter and Buttiker}(2000)}]{Blanter}
\bibinfo{author}{\bibfnamefont{Y.~M.} \bibnamefont{Blanter}} \bibnamefont{and}
  \bibinfo{author}{\bibfnamefont{M.}~\bibnamefont{Buttiker}},
  \bibinfo{journal}{Phys. Rep.} \textbf{\bibinfo{volume}{336}},
  \bibinfo{pages}{1} (\bibinfo{year}{2000}).

\bibitem[{\citenamefont{Korotkov et~al.}(1992)\citenamefont{Korotkov, Averin,
  Likharev, and Vasenko}}]{KorotkovR}
\bibinfo{author}{\bibfnamefont{A.~N.} \bibnamefont{Korotkov}},
  \bibinfo{author}{\bibfnamefont{D.~V.} \bibnamefont{Averin}},
  \bibinfo{author}{\bibfnamefont{K.~K.} \bibnamefont{Likharev}},
  \bibnamefont{and} \bibinfo{author}{\bibfnamefont{S.~A.}
  \bibnamefont{Vasenko}}, in \emph{\bibinfo{booktitle}{Single-Electron
  Tunneling and Mesoscopic Devices}}, edited by
  \bibinfo{editor}{\bibfnamefont{H.}~\bibnamefont{Koch}} \bibnamefont{and}
  \bibinfo{editor}{\bibfnamefont{H.}~\bibnamefont{L{\"u}bbig}}
  (\bibinfo{publisher}{Springer-Verlag}, \bibinfo{address}{Berlin},
  \bibinfo{year}{1992}), vol.~\bibinfo{volume}{31} of
  \emph{\bibinfo{series}{Springer Series in Electronics and Photonics}}, pp.
  \bibinfo{pages}{45--59}.

\bibitem[{\citenamefont{Averin}(2001)}]{Averin}
\bibinfo{author}{\bibfnamefont{D.~V.} \bibnamefont{Averin}}, in
  \emph{\bibinfo{booktitle}{Macroscopic Quantum Coherence and Quantum
  Computing}}, edited by \bibinfo{editor}{\bibfnamefont{D.~V.}
  \bibnamefont{Averin}},
  \bibinfo{editor}{\bibfnamefont{R.}~\bibnamefont{Ruggiero}}, \bibnamefont{and}
  \bibinfo{editor}{\bibfnamefont{P.}~\bibnamefont{Silvestrini}}
  (\bibinfo{publisher}{Kluwer Academic/Plenum Publishers},
  \bibinfo{address}{New York}, \bibinfo{year}{2001}), pp.
  \bibinfo{pages}{399--407}.

\bibitem[{\citenamefont{Johansson et~al.}(2002)\citenamefont{Johansson,
  K{\"a}ck, and Wendin}}]{Johansson}
\bibinfo{author}{\bibfnamefont{G.}~\bibnamefont{Johansson}},
  \bibinfo{author}{\bibfnamefont{A.}~\bibnamefont{K{\"a}ck}}, \bibnamefont{and}
  \bibinfo{author}{\bibfnamefont{G.}~\bibnamefont{Wendin}},
  \bibinfo{journal}{Phys. Rev. Lett.} \textbf{\bibinfo{volume}{88}},
  \bibinfo{pages}{046802} (\bibinfo{year}{2002}).

\bibitem[{\citenamefont{Schoelkopf et~al.}(1998)\citenamefont{Schoelkopf,
  Wahlgren, Kozhevnikov, Delsing, and Prober}}]{Schoelkopf}
\bibinfo{author}{\bibfnamefont{R.~J.} \bibnamefont{Schoelkopf}},
  \bibinfo{author}{\bibfnamefont{P.}~\bibnamefont{Wahlgren}},
  \bibinfo{author}{\bibfnamefont{A.~A.} \bibnamefont{Kozhevnikov}},
  \bibinfo{author}{\bibfnamefont{P.}~\bibnamefont{Delsing}}, \bibnamefont{and}
  \bibinfo{author}{\bibfnamefont{D.~E.} \bibnamefont{Prober}},
  \bibinfo{journal}{Science} \textbf{\bibinfo{volume}{280}},
  \bibinfo{pages}{1238} (\bibinfo{year}{1998}).

\bibitem[{\citenamefont{Hershfield et~al.}(1993)\citenamefont{Hershfield,
  Davies, Hyldgaard, Stanton, and Wilkins}}]{Hershfield3}
\bibinfo{author}{\bibfnamefont{S.}~\bibnamefont{Hershfield}},
  \bibinfo{author}{\bibfnamefont{J.~H.} \bibnamefont{Davies}},
  \bibinfo{author}{\bibfnamefont{P.}~\bibnamefont{Hyldgaard}},
  \bibinfo{author}{\bibfnamefont{C.~J.} \bibnamefont{Stanton}},
  \bibnamefont{and} \bibinfo{author}{\bibfnamefont{J.~W.}
  \bibnamefont{Wilkins}}, \bibinfo{journal}{Phys. Rev. B}
  \textbf{\bibinfo{volume}{47}}, \bibinfo{pages}{1967} (\bibinfo{year}{1993}).

\bibitem[{\citenamefont{Korotkov}(1994)}]{Korotkov1}
\bibinfo{author}{\bibfnamefont{A.~N.} \bibnamefont{Korotkov}},
  \bibinfo{journal}{Phys. Rev. B} \textbf{\bibinfo{volume}{49}},
  \bibinfo{pages}{10381} (\bibinfo{year}{1994}).

\bibitem[{\citenamefont{Korotkov}(1998)}]{Korotkov2}
\bibinfo{author}{\bibfnamefont{A.~N.} \bibnamefont{Korotkov}},
  \bibinfo{journal}{Europhys. Lett.} \textbf{\bibinfo{volume}{43}},
  \bibinfo{pages}{343} (\bibinfo{year}{1998}).

\bibitem[{\citenamefont{van~den Brink}(2002)}]{Brink}
\bibinfo{author}{\bibfnamefont{A.~M.} \bibnamefont{van~den Brink}},
  \bibinfo{journal}{Europhys. Lett.} \textbf{\bibinfo{volume}{58}},
  \bibinfo{pages}{562} (\bibinfo{year}{2002}).

\bibitem[{\citenamefont{Sukhorukov et~al.}(2001)\citenamefont{Sukhorukov,
  Burkard, and Loss}}]{Sukhorukov}
\bibinfo{author}{\bibfnamefont{E.~V.} \bibnamefont{Sukhorukov}},
  \bibinfo{author}{\bibfnamefont{G.}~\bibnamefont{Burkard}}, \bibnamefont{and}
  \bibinfo{author}{\bibfnamefont{D.}~\bibnamefont{Loss}},
  \bibinfo{journal}{Phys. Rev. B} \textbf{\bibinfo{volume}{63}},
  \bibinfo{pages}{125315} (\bibinfo{year}{2001}).

\bibitem[{\citenamefont{Averin and Nazarov}(1992)}]{Averin_Nazarov}
\bibinfo{author}{\bibfnamefont{D.~V.} \bibnamefont{Averin}} \bibnamefont{and}
  \bibinfo{author}{\bibfnamefont{Y.~V.} \bibnamefont{Nazarov}}, in
  \cite{Sin-Cha-Tun}, chap.~\bibinfo{chapter}{6}, p. \bibinfo{pages}{217}.

\bibitem[{\citenamefont{Langreth}(1976)}]{Langreth}
\bibinfo{author}{\bibfnamefont{D.~C.} \bibnamefont{Langreth}}, in
  \emph{\bibinfo{booktitle}{Linear and Nonlinear Transport in solids}}, edited
  by \bibinfo{editor}{\bibfnamefont{J.~T.} \bibnamefont{Devreese}}
  \bibnamefont{and} \bibinfo{editor}{\bibfnamefont{V.~E.} \bibnamefont{van
  Doren}} (\bibinfo{publisher}{Plenum Press}, \bibinfo{address}{New York},
  \bibinfo{year}{1976}), vol.~\bibinfo{volume}{17} of
  \emph{\bibinfo{series}{NATO ASI, Series B: Physics}}.

\bibitem[{\citenamefont{Rammer and Smith}(1986)}]{Rammer}
\bibinfo{author}{\bibfnamefont{J.}~\bibnamefont{Rammer}} \bibnamefont{and}
  \bibinfo{author}{\bibfnamefont{H.}~\bibnamefont{Smith}},
  \bibinfo{journal}{Rev. Mod. Phys.} \textbf{\bibinfo{volume}{58}},
  \bibinfo{pages}{323} (\bibinfo{year}{1986}).

\bibitem[{\citenamefont{Chou et~al.}(1985)\citenamefont{Chou, Su, Hao, and
  Yu}}]{Chou}
\bibinfo{author}{\bibfnamefont{K.-C.} \bibnamefont{Chou}},
  \bibinfo{author}{\bibfnamefont{Z.-B.} \bibnamefont{Su}},
  \bibinfo{author}{\bibfnamefont{B.-L.} \bibnamefont{Hao}}, \bibnamefont{and}
  \bibinfo{author}{\bibfnamefont{L.}~\bibnamefont{Yu}}, \bibinfo{journal}{Phys.
  Rep.} \textbf{\bibinfo{volume}{118}}, \bibinfo{pages}{1}
  (\bibinfo{year}{1985}).

\bibitem[{\citenamefont{K{\"o}nig et~al.}(1996)\citenamefont{K{\"o}nig, Schmid,
  Schoeller, and Sch{\"o}n}}]{Konig1}
\bibinfo{author}{\bibfnamefont{J.}~\bibnamefont{K{\"o}nig}},
  \bibinfo{author}{\bibfnamefont{J.}~\bibnamefont{Schmid}},
  \bibinfo{author}{\bibfnamefont{H.}~\bibnamefont{Schoeller}},
  \bibnamefont{and}
  \bibinfo{author}{\bibfnamefont{G.}~\bibnamefont{Sch{\"o}n}},
  \bibinfo{journal}{Phys. Rev. B} \textbf{\bibinfo{volume}{54}},
  \bibinfo{pages}{16820} (\bibinfo{year}{1996}).

\bibitem[{\citenamefont{Isawa and Horii}(2000)}]{Isawa}
\bibinfo{author}{\bibfnamefont{Y.}~\bibnamefont{Isawa}} \bibnamefont{and}
  \bibinfo{author}{\bibfnamefont{H.}~\bibnamefont{Horii}}, \bibinfo{journal}{J.
  Phys. Soc. Jpn.} \textbf{\bibinfo{volume}{69}}, \bibinfo{pages}{655}
  (\bibinfo{year}{2000}).

\bibitem[{\citenamefont{Spencer and Doniach}(1967)}]{Spencer}
\bibinfo{author}{\bibfnamefont{H.~J.} \bibnamefont{Spencer}} \bibnamefont{and}
  \bibinfo{author}{\bibfnamefont{S.}~\bibnamefont{Doniach}},
  \bibinfo{journal}{Phys. Rev. Lett.} \textbf{\bibinfo{volume}{18}},
  \bibinfo{pages}{23} (\bibinfo{year}{1967}).

\bibitem[{\citenamefont{Kamenev and Andreev}(1999)}]{Kamenev_1}
\bibinfo{author}{\bibfnamefont{A.}~\bibnamefont{Kamenev}} \bibnamefont{and}
  \bibinfo{author}{\bibfnamefont{A.}~\bibnamefont{Andreev}},
  \bibinfo{journal}{Phys. Rev. B} \textbf{\bibinfo{volume}{60}},
  \bibinfo{pages}{2218} (\bibinfo{year}{1999}).

\bibitem[{\citenamefont{Gutman and Gefen}(2001)}]{Gutman_1}
\bibinfo{author}{\bibfnamefont{D.~B.} \bibnamefont{Gutman}} \bibnamefont{and}
  \bibinfo{author}{\bibfnamefont{Y.}~\bibnamefont{Gefen}},
  \bibinfo{journal}{Phys. Rev. B} \textbf{\bibinfo{volume}{64}},
  \bibinfo{pages}{205317} (\bibinfo{year}{2001}).

\bibitem[{not({\natexlab{a}})}]{note10}
\bibinfo{note}{We adopt the time-path $C$ used in Ref. \cite{Kiselev} rather
  than the Keldysh contour $C_{+}+C_{-}$ in Ref. \cite{Chou}, because the path
  $C$ makes the formulation compact in the case where the initial state is in
  equilibrium.}

\bibitem[{not({\natexlab{b}})}]{note1}
\bibinfo{note}{In Ref. \cite{Chou}, the normalization condition of the
  generating functional is $Z|_{J=0}=1$, viz. $Z$ is normalized by $Z_0$ in
  advance.}

\bibitem[{\citenamefont{Mattis}(1988)}]{Mattis}
\bibinfo{author}{\bibfnamefont{D.~C.} \bibnamefont{Mattis}},
  \emph{\bibinfo{title}{The Theory of Magnetism}}, vol.~\bibinfo{volume}{17} of
  \emph{\bibinfo{series}{Springer Series in Solid-State Sciences}}
  (\bibinfo{publisher}{Springer-Verlag}, \bibinfo{address}{Berlin Heidelberg
  New York}, \bibinfo{year}{1988}), \bibinfo{edition}{2nd} ed.,
  \bibinfo{note}{p. 90}.

\bibitem[{\citenamefont{Babichenko and Kozlov}(1986)}]{Babichenko}
\bibinfo{author}{\bibfnamefont{V.~S.} \bibnamefont{Babichenko}}
  \bibnamefont{and} \bibinfo{author}{\bibfnamefont{A.~N.}
  \bibnamefont{Kozlov}}, \bibinfo{journal}{Solid State Commun.}
  \textbf{\bibinfo{volume}{59}}, \bibinfo{pages}{39} (\bibinfo{year}{1986}).

\bibitem[{not({\natexlab{c}})}]{note7}
\bibinfo{note}{The method to perform the path integral {\it directly} on $C$ is
  demonstrated in an appendix of Refs. \cite{Okumura,Utsumi_3}. A solution of
  differential equations defined on $C$ is shown in an appendix of Ref.
  \cite{Utsumi_3}.}

\bibitem[{\citenamefont{Utsumi et~al.}(2002{\natexlab{a}})\citenamefont{Utsumi,
  Imamura, Hayashi, and Ebisawa}}]{Utsumi_3}
\bibinfo{author}{\bibfnamefont{Y.}~\bibnamefont{Utsumi}},
  \bibinfo{author}{\bibfnamefont{H.}~\bibnamefont{Imamura}},
  \bibinfo{author}{\bibfnamefont{M.}~\bibnamefont{Hayashi}}, \bibnamefont{and}
  \bibinfo{author}{\bibfnamefont{H.}~\bibnamefont{Ebisawa}},
  \bibinfo{journal}{Phys. Rev. B} \textbf{\bibinfo{volume}{66}},
  \bibinfo{pages}{024513} (\bibinfo{year}{2002}{\natexlab{a}}).

\bibitem[{\citenamefont{Zagoskin}(1998)}]{Zagoskin}
\bibinfo{author}{\bibfnamefont{A.~M.} \bibnamefont{Zagoskin}},
  \emph{\bibinfo{title}{Quantum Theory of Many-Body Systems}}
  (\bibinfo{publisher}{Springer-Velag}, \bibinfo{address}{New York},
  \bibinfo{year}{1998}).

\bibitem[{not({\natexlab{d}})}]{note8}
\bibinfo{note}{Our treatment of the finite transport voltage is equivalent to
  the widely used one\cite{Caroli_1,Caroli_2,Caroli_3,Caroli_4,Isawa2,Meir_1},
  where the voltage difference between two leads is included in the initial
  distribution function. Two treatments are related each other by a gauge
  transformation.}

\bibitem[{not({\natexlab{e}})}]{note9}
\bibinfo{note}{There are two differences between our definition and that in
  Ref. \cite{Gutman_1}. First, we use the generating functional for connected
  Green function $W$ rather than $Z$, because we are interested in the
  correlation function of the current (or charge) {\it fluctuation} operator.
  Second, we propose the modified expression for the noise: Besides the
  standard definition, the first term of Eq. (\ref{eqn:SII}) or Eq.
  (\ref{eqn:SQQ}), we introduce the additional second term. It circumvents the
  uncertainty related to the order of operators (Appendix.
  \ref{appendix:expressions}).}

\bibitem[{\citenamefont{Hershfield et~al.}(1992)\citenamefont{Hershfield,
  Davies, and Wilkins}}]{Hershfield1}
\bibinfo{author}{\bibfnamefont{S.}~\bibnamefont{Hershfield}},
  \bibinfo{author}{\bibfnamefont{J.~H.} \bibnamefont{Davies}},
  \bibnamefont{and} \bibinfo{author}{\bibfnamefont{J.~W.}
  \bibnamefont{Wilkins}}, \bibinfo{journal}{Phys. Rev. B}
  \textbf{\bibinfo{volume}{46}}, \bibinfo{pages}{7046} (\bibinfo{year}{1992}).

\bibitem[{\citenamefont{Hershfield}(1992)}]{Hershfield2}
\bibinfo{author}{\bibfnamefont{S.}~\bibnamefont{Hershfield}},
  \bibinfo{journal}{Phys. Rev. B} \textbf{\bibinfo{volume}{46}},
  \bibinfo{pages}{7061} (\bibinfo{year}{1992}).

\bibitem[{not({\natexlab{f}})}]{note5}
\bibinfo{note}{Our notations, $\alpha^K_{\rm r}(\varepsilon)$ and
  $G_c^C(\varepsilon)$, correspond to $-2 \ri \pi \alpha_{\rm r}(\varepsilon)$
  and $-2 \ri \pi A(\varepsilon)$ in Ref. \cite{Schoeller_Schon},
  respectively.}

\bibitem[{\citenamefont{Utsumi et~al.}(2002{\natexlab{b}})\citenamefont{Utsumi,
  Imamura, Hayashi, and Ebisawa}}]{Utsumi_4}
\bibinfo{author}{\bibfnamefont{Y.}~\bibnamefont{Utsumi}},
  \bibinfo{author}{\bibfnamefont{H.}~\bibnamefont{Imamura}},
  \bibinfo{author}{\bibfnamefont{M.}~\bibnamefont{Hayashi}}, \bibnamefont{and}
  \bibinfo{author}{\bibfnamefont{H.}~\bibnamefont{Ebisawa}},
  \bibinfo{journal}{Physica C} \textbf{\bibinfo{volume}{367}},
  \bibinfo{pages}{237} (\bibinfo{year}{2002}{\natexlab{b}}).

\bibitem[{\citenamefont{Lesovik}(1989)}]{Lesovik}
\bibinfo{author}{\bibfnamefont{G.~B.} \bibnamefont{Lesovik}},
  \bibinfo{journal}{JETP Lett.} \textbf{\bibinfo{volume}{49}},
  \bibinfo{pages}{592} (\bibinfo{year}{1989}).

\bibitem[{not({\natexlab{g}})}]{note4}
\bibinfo{note}{Equation (42) in Ref. \cite{Korotkov1} is the corresponding
  result (for $T \ll E_C$). We adopt different notations and definitions; $S_{Q
  Q}(0)$ corresponds to $S_{\varphi \, \varphi}(0) \, (2/C^2)$. $\Gamma_{\rm
  L(R) \, I}$ and $\Gamma_{\rm I \, L(R)}$ correspond to $\Gamma_{1(2)}^{+}$
  and $\Gamma_{1(2)}^{-}$, respectively.}

\bibitem[{not({\natexlab{h}})}]{note11}
\bibinfo{note}{We confirmed that following numerical calculations achieve the
  spectral sum rule for $c$-field GF, $\int \rd \varepsilon \, G^C_c=-2 \ri
  \pi$, up to $0.005\%$.}

\bibitem[{\citenamefont{Shimizu and Ueda}(1992)}]{Shimizu}
\bibinfo{author}{\bibfnamefont{A.}~\bibnamefont{Shimizu}} \bibnamefont{and}
  \bibinfo{author}{\bibfnamefont{M.}~\bibnamefont{Ueda}},
  \bibinfo{journal}{Phys. Rev. Lett.} \textbf{\bibinfo{volume}{69}},
  \bibinfo{pages}{1403} (\bibinfo{year}{1992}).

\bibitem[{\citenamefont{Ueda and Shimizu}(1993)}]{Ueda}
\bibinfo{author}{\bibfnamefont{M.}~\bibnamefont{Ueda}} \bibnamefont{and}
  \bibinfo{author}{\bibfnamefont{A.}~\bibnamefont{Shimizu}},
  \bibinfo{journal}{J. Phys. Soc. Jpn.} \textbf{\bibinfo{volume}{62}},
  \bibinfo{pages}{2994} (\bibinfo{year}{1993}).

\bibitem[{\citenamefont{Khlus}(1987)}]{Khlus}
\bibinfo{author}{\bibfnamefont{V.~A.} \bibnamefont{Khlus}},
  \bibinfo{journal}{Sov. Phys. JETP} \textbf{\bibinfo{volume}{66}},
  \bibinfo{pages}{1243} (\bibinfo{year}{1987}).

\bibitem[{\citenamefont{Utsumi et~al.}(cond-mat/0205114)\citenamefont{Utsumi,
  Imamura, Hayashi, and Ebisawa}}]{Utsumi_5}
\bibinfo{author}{\bibfnamefont{Y.}~\bibnamefont{Utsumi}},
  \bibinfo{author}{\bibfnamefont{H.}~\bibnamefont{Imamura}},
  \bibinfo{author}{\bibfnamefont{M.}~\bibnamefont{Hayashi}}, \bibnamefont{and}
  \bibinfo{author}{\bibfnamefont{H.}~\bibnamefont{Ebisawa}},
  \bibinfo{journal}{to appear in proceedings of {\it The International
  Symposium on Mesoscopic Superconductivity and Spintronics (MS+S2002)}}
  (\bibinfo{year}{cond-mat/0205114}).

\bibitem[{\citenamefont{Utsumi et~al.}(cond-mat/0211036)\citenamefont{Utsumi,
  Imamura, Hayashi, and Ebisawa}}]{Utsumi_6}
\bibinfo{author}{\bibfnamefont{Y.}~\bibnamefont{Utsumi}},
  \bibinfo{author}{\bibfnamefont{H.}~\bibnamefont{Imamura}},
  \bibinfo{author}{\bibfnamefont{M.}~\bibnamefont{Hayashi}}, \bibnamefont{and}
  \bibinfo{author}{\bibfnamefont{H.}~\bibnamefont{Ebisawa}},
  \bibinfo{journal}{to appear in proceedings of {\it The 23rd International
  Conference of Low Temperature Physics (August 20-27, 2002, Hiroshima) }}
  (\bibinfo{year}{cond-mat/0211036}).

\bibitem[{\citenamefont{Belzig and Nazarov}(2001)}]{Belzig1}
\bibinfo{author}{\bibfnamefont{W.}~\bibnamefont{Belzig}} \bibnamefont{and}
  \bibinfo{author}{\bibfnamefont{Y.~V.} \bibnamefont{Nazarov}},
  \bibinfo{journal}{Phys. Rev. Lett.} \textbf{\bibinfo{volume}{87}},
  \bibinfo{pages}{067006} (\bibinfo{year}{2001}).

\bibitem[{\citenamefont{Bagrets and Nazarov}(unpublished)}]{Bagrets}
\bibinfo{author}{\bibfnamefont{D.~A.} \bibnamefont{Bagrets}} \bibnamefont{and}
  \bibinfo{author}{\bibfnamefont{Y.~V.} \bibnamefont{Nazarov}},
  \bibinfo{journal}{cond-mat/0207624}  (\bibinfo{year}{unpublished}).

\bibitem[{not({\natexlab{i}})}]{note3}
\bibinfo{note}{We can confirm the term ${\rm Tr} [\mtau^1 \tilde{\Sigma}_{\rm
  r} \mtau^1 \tilde{g}_c ]=\Sigma_{\rm r}^A g_c^A+\Sigma_{\rm r}^R g_c^R $ is 0
  from the analytic properties of the step function $\int \rd \varepsilon'
  1/((\varepsilon'+\varepsilon+\ri \eta)(\varepsilon+\ri \eta))=0$, viz.
  $\theta(t) \theta(-t)=0$. The standard definition for noise, the first term
  of Eq. (\ref{eqn:SII}) or Eq. (\ref{eqn:SQQ}), includes such terms, and thus
  one must remove them carefully. Our definition does not include such terms,
  which simplifies practical calculations for the second moment.}

\bibitem[{\citenamefont{Grabert and Devoret}(1992)}]{Sin-Cha-Tun}
\bibinfo{editor}{\bibfnamefont{H.}~\bibnamefont{Grabert}} \bibnamefont{and}
  \bibinfo{editor}{\bibfnamefont{M.~H.} \bibnamefont{Devoret}}, eds.,
  \emph{\bibinfo{title}{Single Charge Tunneling}}, vol. \bibinfo{volume}{294}
  of \emph{\bibinfo{series}{NATO ASI Series B}} (\bibinfo{publisher}{Plenum
  Press}, \bibinfo{address}{New York and London}, \bibinfo{year}{1992}).

\bibitem[{\citenamefont{Kiselev and Oppermann}(2000)}]{Kiselev}
\bibinfo{author}{\bibfnamefont{M.~N.} \bibnamefont{Kiselev}} \bibnamefont{and}
  \bibinfo{author}{\bibfnamefont{R.}~\bibnamefont{Oppermann}},
  \bibinfo{journal}{Phys. Rev. Lett.} \textbf{\bibinfo{volume}{85}},
  \bibinfo{pages}{5631} (\bibinfo{year}{2000}).

\bibitem[{\citenamefont{Okumura and Tanimura}(1996)}]{Okumura}
\bibinfo{author}{\bibfnamefont{K.}~\bibnamefont{Okumura}} \bibnamefont{and}
  \bibinfo{author}{\bibfnamefont{Y.}~\bibnamefont{Tanimura}},
  \bibinfo{journal}{Phys. Rev. E} \textbf{\bibinfo{volume}{53}},
  \bibinfo{pages}{214} (\bibinfo{year}{1996}).

\bibitem[{\citenamefont{Caroli et~al.}(1971{\natexlab{a}})\citenamefont{Caroli,
  Combescot, Nozieres, and Saint-James}}]{Caroli_1}
\bibinfo{author}{\bibfnamefont{C.}~\bibnamefont{Caroli}},
  \bibinfo{author}{\bibfnamefont{R.}~\bibnamefont{Combescot}},
  \bibinfo{author}{\bibfnamefont{P.}~\bibnamefont{Nozieres}}, \bibnamefont{and}
  \bibinfo{author}{\bibfnamefont{D.}~\bibnamefont{Saint-James}},
  \bibinfo{journal}{J. Phys. {\bf C4}} p. \bibinfo{pages}{916}
  (\bibinfo{year}{1971}{\natexlab{a}}).

\bibitem[{\citenamefont{Caroli et~al.}(1971{\natexlab{b}})\citenamefont{Caroli,
  Combescot, Lederer, Nozieres, and Saint-James}}]{Caroli_2}
\bibinfo{author}{\bibfnamefont{C.}~\bibnamefont{Caroli}},
  \bibinfo{author}{\bibfnamefont{R.}~\bibnamefont{Combescot}},
  \bibinfo{author}{\bibfnamefont{D.}~\bibnamefont{Lederer}},
  \bibinfo{author}{\bibfnamefont{P.}~\bibnamefont{Nozieres}}, \bibnamefont{and}
  \bibinfo{author}{\bibfnamefont{D.}~\bibnamefont{Saint-James}},
  \bibinfo{journal}{J. Phys. {\bf C4}} p. \bibinfo{pages}{2598}
  (\bibinfo{year}{1971}{\natexlab{b}}).

\bibitem[{\citenamefont{Combescot}(1971)}]{Caroli_3}
\bibinfo{author}{\bibfnamefont{R.}~\bibnamefont{Combescot}},
  \bibinfo{journal}{J. Phys. {\bf C4}} p. \bibinfo{pages}{2611}
  (\bibinfo{year}{1971}).

\bibitem[{\citenamefont{Caroli et~al.}(1972)\citenamefont{Caroli, Combescot,
  Nozieres, and Saint-James}}]{Caroli_4}
\bibinfo{author}{\bibfnamefont{C.}~\bibnamefont{Caroli}},
  \bibinfo{author}{\bibfnamefont{R.}~\bibnamefont{Combescot}},
  \bibinfo{author}{\bibfnamefont{P.}~\bibnamefont{Nozieres}}, \bibnamefont{and}
  \bibinfo{author}{\bibfnamefont{D.}~\bibnamefont{Saint-James}},
  \bibinfo{journal}{J. Phys. {\bf C5}} p.~\bibinfo{pages}{21}
  (\bibinfo{year}{1972}).

\bibitem[{\citenamefont{Isawa and Kanechika}(1991)}]{Isawa2}
\bibinfo{author}{\bibfnamefont{Y.}~\bibnamefont{Isawa}} \bibnamefont{and}
  \bibinfo{author}{\bibfnamefont{M.}~\bibnamefont{Kanechika}},
  \bibinfo{journal}{J. Phys. Soc. Jpn.} \textbf{\bibinfo{volume}{60}},
  \bibinfo{pages}{4013} (\bibinfo{year}{1991}).

\bibitem[{\citenamefont{Meir and Wingreen}(1992)}]{Meir_1}
\bibinfo{author}{\bibfnamefont{Y.}~\bibnamefont{Meir}} \bibnamefont{and}
  \bibinfo{author}{\bibfnamefont{N.~S.} \bibnamefont{Wingreen}},
  \bibinfo{journal}{Phys. Rev. Lett.} \textbf{\bibinfo{volume}{68}},
  \bibinfo{pages}{2512} (\bibinfo{year}{1992}).

\end{thebibliography}

\end{document}